\documentclass[aps,11pt,preprintnumbers,nofootinbib,floatfix]{revtex4}

\usepackage{graphicx}
\usepackage{epsfig}
\usepackage{dcolumn}
\usepackage{subfigure}
\usepackage{multirow}
\usepackage{amsmath}
\usepackage{amssymb}

\begin{document}

\title{A more realistic holographic model of color superconductivity with the higher derivative corrections}


\author{Cao H. Nam}
\email{nam.caohoang@phenikaa-uni.edu.vn}  
\affiliation{Phenikaa Institute for Advanced Study and Faculty of Fundamental Sciences, Phenikaa University, Yen Nghia, Ha Dong, Hanoi 12116, Vietnam}
\date{\today}

\begin{abstract}%
In this paper, we have constructed a bottom-up holographic model for the color superconductivity (CSC) of the Yang-Mills theory with including the higher derivative corrections which allow to study the CSC phase with the color number $N_c\geq2$. First, we consider the CSC phase transition in the context of Einstein-Gauss-Bonnet (EGB) gravity. We analyze the Cooper pair condensate in the deconfinement and confinement phases which are dual to the planar GB-RN-AdS black hole and GB-AdS soliton, respectively, where the backreaction of the matter part is taken into account. By examining the breakdown of the Breitenlohner-Freedman bound in the background of the planar GB-RN-AdS black hole, we find that the positive GB coupling parameter $\alpha>0$ leads to a lower upper bound of the color number in comparison to Einstein gravity where the CSC phase for $N_c\geq2$ is not realized. But, with the $\alpha<0$ case it is possible to observe the Cooper pair condensate for $N_c\geq2$ with the reasonable magnitude of $\alpha$. This is confirmed and the corresponding phase diagram is found by solving numerically the equations of motion for the gravitational system. In addition, we show that the CSC phase disappears in the confinement phase for the magnitude of $\alpha$ below a certain value which means that beyond that value it might lead to the breakdown region of the EGB gravity in investigating the CSC phase. However, the CSC phase transition occurring with $N_c\geq2$ requires the magnitude of the GB coupling parameter to be rather large. As a result, the GB term would no longer be considered as the correction and it also violates the boundary causality bound. We resolve this problem by including additionally the higher derivative correction for the Maxwell electrodynamics and the non-minimal coupled Maxwell field.
\end{abstract}

\maketitle

\section{Introduction}
It is expected in quantum chromodynamics (QCD) that at sufficiently high chemical potential (density) and low temperature quarks condense into Cooper pairs in analogy to the condensation of electrons in the conventional metallic superconductors \cite{Alford2008}. Unlike the condensation of the electron pairs where the Coulomb interaction between them is repulsive and has to be overcome by an attraction caused by the coupling between electrons with phonons, the strong interaction between two quarks is attractive (in the color-antisymmetric channel) and thus the Bardeen-Cooper-Schrieffer (BCS) mechanism applied to the quark pairs is more direct than its original setting. The quark pairs carry the net color charge or in other words they are gauge non-invariant operators.\footnote{It was indicated that the gauge non-invariant operators constructed by the quark pairs are suppressed in the limit of large color number \cite{Deryagin1992,Son2000}, which is considered as one of the large obstacles for the relevant investigations.} Therefore, the condensation of the quark pairs breaks spontaneously the $SU(3)_C$ gauge symmetry of QCD and gives rise the masses for the gluon via Higgs mechanism. This phenomena is thus referred to the color superconductivity (CSC). It is interesting to study the CSC phase from both theoretical and phenomenological aspects. The quark pairs have color and flavor degrees of freedom besides the spin one and hence there are different condensation patterns of which the color-flavor locked phase \cite{Alford1999} is well-known. Also, the CSC phase might occur in the cores of neutron stars with the densities possibly reaching up ten times nuclear-matter saturation density.

At very large temperature or chemical potential, QCD becomes weakly coupled due to the asymptotic freedom and hence an exactly analytic study of the quark matter is possible. However, in the nonperturbative region the investigations are mainly based the phenomenological models at which many important features are missed. Many nonperturbative investigations can be performed by using the numerical simulation, however it can be inaccessible at finite chemical potential due to the sign problem of the Euclidean action. 

Another approach for investigating the properties of the strongly coupled theories at finite temperature and chemical potential is via the AdS/CFT correspondence \cite{Maldacena,Witten,Gubser} which relates the weakly coupled gravitational theory in $d$-dimensional AdS spacetime and the strongly coupled conformal theory of $d-1$ dimensions living at the boundary of AdS spacetime, referred to the holographic approach. The  application of the holographic approach to QCD has been investigated with the top-down approach at which the holographic QCD models arise directly from the ten-dimensional superstring theory
\cite{Karch2002,Sakai2005,Erdmenger2007,Horigome2007,Erdmenger2008,Chen2010,Dasgupta2019,
Misra2020}. From the bottom-up approach motivated by the phenomenological reasons, the holographic QCD models were introduced with ignoring the backreaction on the spacetime geometry background \cite{Brodsky2004,Erlich2005,Karch2006,Miranda2009,Braga2016,Braga2018,Mamani2019} and with considering the backreaction \cite{Csaki2007,Batell2008,Dudal2017,HMamani2019,Chen2020}.

Motivated by the application of the holographic approach in condensed matter \cite{Hartnoll2008,Herzog2008}, the holographic model for the CSC phase transition has been constructed in the bottom-up approach \cite{Basu2011,Fadafan2018,Ghoroku2019}. In Ref. \cite{Basu2011}, the bulk system consists of Einstein gravity coupled to a $U(1)$ gauge field and a real scalar field in six dimensions with the boundary geometry $R^{3,1}\times S^1$. The scalar field tends to condense in the near-horizon region at a very low temperature and the CSC phase transition is found due to the fact that the near-horizon geometry of planar Reissner-Nordstrom (RN) AdS black hole is $AdS_2\times R^4$ \cite{Herzog2008,Iqbal2010} corresponding to the new instability bound. However, the real scalar field does not correspond to a diquark operator. Therefore, Ref. \cite{Fadafan2018} considered the complex scalar field (rather than the real one) whose $U(1)$ charge is regarded as the baryon number of the diquark operator. In that work, the backreaction of the matter fields is ignored and the CSC phase is found to appear above a critical chemical potential. In particular, a detail and profound investigation about the CSC phase transition in the Yang-Mills (YM) theory was performed in Ref. \cite{Ghoroku2019} where the authors investigated the CSC phase transition for both the deconfinement and the confinement phases with including the backreation of the matter part. The authors indicated that there is the CSC phase transition in the decconfinement phase but not in the confinement phase for the color number $N_c=1$. But, for $N_c\geq2$ this phase transition does not appear in both the decconfinement and the confinement phases, which is thus unrealistic as the YM theory.

The aim of the present paper is to extend the work in Ref. \cite{Ghoroku2019} to get the CSC phase transition with the color number $N_c\geq2$. In order to do that, first we instead construct a gravitational dual model in Einstein-Gauss-Bonnet (EGB) gravity which is an extension of Einstein gravity with including the higher curvature corrections written as the Gauss-Bonnet (GB) term. We analyze the CSC phase transition in both the decconfinement and the confinement phases which are dual to the planar GB-RN-AdS black hole and GB-AdS soliton, respectively, and indicate the role of the GB term on the occurrence of the CSC phase for $N_c\geq2$. We indicate that the CSC phase transition with $N_c\geq2$ requires the magnitude of the GB coupling parameter to be rather large, which is thus beyond the region of the classical gravity and violates the boundary causality bound. In order to resolve this problem, we consider additionally the higher derivative corrections which come from the matter fields and the non-minimal coupled Maxwell field. We find that there actually exists the small values for the parameters associated with the higher derivative corrections where they work to realize the CSC phase transition with $N_c\geq2$.

The organization of the paper is as follows. In Sec. \ref{model}, we introduce the gravitational dual model in the context of the EGB gravity to investigate the CSC phase transition where the backreaction of the matter part is taken into account. In Sec. \ref{HCSC}, we study the CSC phase transition for $N_c\geq2$ in the decconfinement and confinement phases and look at the role of the GB term on this phase transition. In Sec. \ref{HCSC2}, we study the effects of the higher derivative corrections coming from the matter fields and the non-minimal coupled Maxwell field on the CSC phase transition for $N_c\geq2$. Finally, we conclude our main results in Sec. \ref{conclu}.

\section{\label{model} Model setup}

In this section, we introduce the gravitational dual model in the framework of the six-dimensional EGB gravity for the CSC phase transition, given by the following action
\begin{equation}
S_{\text{bulk}}=\frac{1}{2\kappa^2_6}\int
d^6x\sqrt{-g}\left[R-2\Lambda+\widetilde{\alpha}\mathcal{L}_{GB}+\mathcal{L}_{\text{mat}}\right],\label{EGB-ED-adS}
\end{equation}
where $\Lambda$ is the cosmological constant defined in terms of the asymptotic AdS radius $l$ as $\Lambda=-\frac{10}{l^2}$, $\mathcal{L}_{GB}$ is the GB term given by
\begin{equation}
  \mathcal{L}_{GB}=R^2-4R_{\mu\nu}R^{\mu\nu}+R_{\mu\nu\rho\lambda}R^{\mu\nu\rho\lambda},
\end{equation}
$\widetilde{\alpha}$ is the GB coupling parameter,\footnote{The GB term can be naturally obtained from the low-energy limit of heterotic string theory \cite{Zwiebach1985,Witten1986,Gross1987,Tseytlin1987,Bento1996} where the GB coupling parameter $\widetilde{\alpha}$ is regarded as the inverse string tension.} and $\mathcal{L}_{\text{mat}}$ is the matter Lagrangian. In the bottom-up construction, the matter Lagrangian for the holographic model which consists of a $U(1)$ gauge field $A_\mu$ and a complex scalar field $\psi$ is given as
\begin{eqnarray}
\mathcal{L}_{\text{mat}}=-\frac{1}{4}F_{\mu\nu}F^{\mu\nu}-|(\nabla_\mu-iqA_\mu)\psi|^2-m^2|\psi|^2.
\end{eqnarray}
In this Lagrangian, the $U(1)$ gauge field is regarded as the dual description of the current of the baryon number whose time component describes the baryon charge density and the chemical potential of the quarks. Whereas, the complex scalar field $\psi$ is dual to the diquark operator in the boundary field theory and $q$ is its $U(1)$ charge which is regarded as the baryon number of the diquark operator. Note that, the baryon number of the diquark operator is related the color number $N_c$ as $q=\frac{2}{N_c}$. In the following, we set $1/2\kappa^2_6=1$ and $l=1$.

Before proceeding, let us pause here to discuss whether the terms with the further powers of the curvature tensors such as $R^3$ or $R^4$ would play the important role in the present holographic model. In general, the action for gravity in six dimensions which includes the higher derivative corrections is written as follows
\begin{eqnarray}
S_{\text{gra}}=\frac{1}{2\kappa^2_6}\int
d^6x\sqrt{-g}\left[R+\frac{20}{l^2}+l^2\left(\widetilde{\alpha}_1R^2+\widetilde{\alpha}_2R_{\mu\nu}R^{\mu\nu}+\widetilde{\alpha}_3R_{\mu\nu\rho\lambda}R^{\mu\nu\rho\lambda}\right)+\cdots\right],
\end{eqnarray} 
where the ellipse refers to the terms with the further powers of the curvature tensors, the couplings associated with the curvature-squared terms have been parameterized with the asymptotic AdS radius $l$, and $\widetilde{\alpha}_i\sim\ell^2_P/l^2$ are the dimensionless couplings with $\ell_P$ to be the Planck length in six dimensions. Note that, the dimensionless couplings corresponding to the six- and further derivative terms are proportional to $\ell^4_P/l^4$ and the further powers of $\ell^2_P/l^2$, respectively. For the region of the classical gravity, the asymptotic AdS radius $l$ is much larger than the Planck length $\ell_P$, i.e. $\ell^2_P/l^2\ll1$. Since we have $\widetilde{\alpha}_i\sim\ell^2_P/l^2\ll1$ for the four-derivative (curvature-squared) terms. Compared to the four-derivative terms, the further derivative terms are more strongly suppressed by the further powers of $\ell^2_P/l^2$. For instance, the dimensionless couplings $\lambda_i$ and $f_i$ which correspond to the six- and eight-derivative terms are proportional to $\ell^4_P/l^4$ and $\ell^6_P/l^6$, respectively. This suggests $f_i\ll\lambda_i\ll\widetilde{\alpha}_i$. Therefore, in the region of the classical gravity, the contributions coming from the terms with the further powers of the curvature tensors such as $R^3$ or $R^4$ are small compared to the curvature-squared terms and hence they can be left. Otherwise, once the curvature-squared terms become important, i.e. $\widetilde{\alpha}_i\sim 1$, the further powers of the curvature tensors such as $R^3$ or $R^4$ would no longer be ignored and thus it is beyond the region of the classical gravity.

It was pointed out in \cite{Camanho2010,Escobedo2010}, there are the constraints imposed by the causality of the boundary field theory as, $-51/196\leq\alpha\leq39/256$, where $\alpha\equiv6\widetilde{\alpha}$. As indicated later by Hofman \cite{Hofman2009}, the bounds obtained from the causality constraints of the boundary field theory should not be a feature of the thermal CFTs but the causality violation reflects a fact that the interaction can occur in the asymptotic region close to the boundary. Also, for the better understanding of the effects of the GB term on the CSC phase transition, in this section we permit the following range of the GB coupling parameter, $\alpha\in(-\infty,1/4]$, where the upper bound is imposed to avoid a naked singularity in the pure GB-AdS solution.

Varying the action (\ref{EGB-ED-adS}) with respect to the metric, vector, and scalar fields, we obtain the equations of motion as
\begin{eqnarray}
G_{\mu\nu}+\widetilde{\alpha}H_{\mu\nu}-\frac{10}{l^2}g_{\mu\nu}&=&T_{\mu\nu},\nonumber\\
\nabla_\mu F^{\mu\nu}-iq\left[\psi^*(\nabla^\nu-iqA^\nu)\psi-\psi(\nabla^\nu+iqA^\nu)\psi^*\right]&=&0,\nonumber\\
(\nabla_\mu-iqA_\mu)(\nabla^\mu-iqA^\mu)\psi-m^2\psi&=&0,\label{EOM}
\end{eqnarray}
where
\begin{eqnarray}
H_{\mu\nu}&=&2\left(RR_{\mu\nu}-2R_{\mu\sigma}{R^\sigma}_\nu-2R_{\mu\sigma\nu\rho}R^{\sigma\rho}+{R_\mu}^{\rho\sigma\lambda}R_{\nu\rho\sigma\lambda}\right)-\frac{1}{2}g_{\mu\nu}\mathcal{L}_{GB},\nonumber\\
T_{\mu\nu}&=&\frac{1}{2}F_{\mu\lambda}{F_\nu}^\lambda+\frac{1}{2}\left[(\nabla_\nu-iqA_\nu)\psi(\nabla_\mu+iqA_\mu)\psi^*+\mu\leftrightarrow\nu\right]+\frac{1}{2}g_{\mu\nu}\mathcal{L}_{\text{mat}}.
\end{eqnarray}
In order to solve these equations of motion, we first need to take the ansatz for the metric, vector, and scalar fields. We are interested in two solutions of the first equation of Eq. (\ref{EOM}) which are dual to the deconfinement and confinement phases in the boundary field theory. More specifically, the ansatz for the metric field is given by Eqs. (\ref{BHa}) and (\ref{AdSs}) corresponding to the deconfinement and confinement phases, respectively. 
The ansatz for the vector and scalar fields read
\begin{eqnarray}
A_\mu dx^\mu=\phi(r)dt,\ \ \ \ \psi=\psi(r).\label{vec-scal-ans}
\end{eqnarray}
For each of the deconfinement and confinement phases, we study the CSC phase transition by solving Eq. (\ref{EOM}) with the suitable boundary conditions in order to find the configuration with or without nontrivial scalar which the value of the scalar field is nonzero. 

The CSC phase appears due to the condensation of the scalar field corresponding to the spontaneously broken $U(1)$ symmetry. In the canonical ensemble where the charge is kept fixed, the condensation of the scalar field is triggered by the chemical potential associated with the quark number density. Near the critical chemical potential, the value of the scalar field approaches zero and since the backreaction of the scalar field on the spacetime metric is negligible. On the other hand, the backreaction of the matter on the spacetime metric in this situation only comes from the vector field. 

The spacetime geometry dual to the deconfinement phase is the planar black hole solution whose line element is given by the following ansatz
\begin{eqnarray}
ds^2=r^2\left(-f(r)dt^2+h_{ij}dx^idx^j+dy^2\right)+\frac{dr^2}{r^2f(r)},\label{BHa}
\end{eqnarray}
where $h_{ij}dx^idx^j=dx^2_1+dx^2_2+dx^2_3$ is the line element of the $3$-dimensional planar hypersurface, and the direction $y$ is compacted with the circle radius $R_y$. The event horizon radius $r_+$ satisfies $f(r_+)=0$. The temperature of the boundary field theory is identified as the Hawking temperature as, $T=\frac{r^2_+f'(r_+)}{4\pi}$. In this configuration of the spacetime geometry, we find the equations for $f(r)$, $\phi(r)$, and $\psi(r)$ from Eq. (\ref{EOM}) as
\begin{eqnarray}
\alpha\left[2f'(r)r+5f(r)\right]f(r)-rf'(r)-5f(r)+\frac{5}{l^2}&=&\frac{1}{8}\phi'(r)^2,\label{r-f-Eq}\\
\phi''(r)+\frac{4}{r}\phi'(r)-\frac{2q^2\psi^2(r)}{r^2f(r)}\phi(r)&=&0,\label{r-phi-Eq}\\
\psi''(r)+\left[\frac{f'(r)}{f(r)}+\frac{6}{r}\right]\psi'(r)+\frac{1}{r^2f(r)}\left[\frac{q^2\phi^2(r)}{r^2f(r)}-m^2\right]\psi(r)&=&0.\label{r-psi-Eq}
\end{eqnarray}
Near the AdS boundary ($r\rightarrow\infty$), the spacetime geometry becomes the planar GB-RN-AdS black hole with $f(r)$ given as
\begin{eqnarray}
f(r)=\frac{1}{2\alpha}\left[1-\sqrt{1-4\alpha\left(1-\frac{r^5_+}{r^5}\right)+\frac{3\alpha\mu^2}{2r^2_+}\left(\frac{r_+}{r}\right)^5\left(1-\frac{r^3_+}{r^3}\right)}\right].\label{GBRNAdSfr}
\end{eqnarray}
Whereas, the asymptotic behavior of the matter fields are given by
\begin{eqnarray}
\phi(r)&=&\mu-\frac{\bar{d}}{r^3},\nonumber\\
\psi(r)&=&\frac{J_C}{r^{\Delta_-}}+\frac{C}{r^{\Delta_+}},\label{phi-psi-asy-beh}
\end{eqnarray}
where $\mu$, $\bar{d}$, $J_C$, and $C$ are regarded as the chemical potential, charge density, source, and the condensate value (VEV) of the diquark operator dual to $\psi$, respectively, and the conformal dimensions $\Delta_{\pm}$ read
\begin{eqnarray}
\Delta_\pm=\frac{1}{2}\left(5\pm\sqrt{25+4m^2l^2_{\text{eff}}}\right),\ \ \ \ l^2_{\text{eff}}=\frac{2\alpha}{1-\sqrt{1-4\alpha}},
\end{eqnarray}
which suggests the Breitenlohner-Freedman (BF) bound \cite{Freedman1982,Breitenlohner1982} as
\begin{equation}
m^2l^2_{\text{eff}}\geq-\frac{25}{4}.
\end{equation}
Because the scalar field $\psi$ is dual to the quark pair, the conformal dimension $\Delta_+$ of $C$ should be $\Delta_+=2\times\frac{d-2}{2}$ which is equal to four for the case of $d=6$. This suggests $m^2l^2_{\text{eff}}=-4$ and thus $\Delta_-=1$. Near the event horizon, the solution must have the following expansions
\begin{eqnarray}
f(r)&=&f_0+f_1(r-r_+)+f_2(r-r_+)^2+f_3(r-r_+)^3+\cdots,\nonumber\\
\phi(r)&=&\phi_0+\phi_1(r-r_+)+\phi_2(r-r_+)^2+\phi_3(r-r_+)^3+\cdots,\nonumber\\
\psi(r)&=&\psi_0+\psi_1(r-r_+)+\psi_2(r-r_+)^2+\psi_3(r-r_+)^3+\cdots,
\end{eqnarray}
where $f_i$, $\phi_i$, and $\psi_i$ (with $i=0, 1, 2,\cdots$) are constants. Because the function $f(r)$ vanishes at the event horizon, we find $f_0=0$. Furthermore, we need to impose the regularity condition for the matter fields at the event horizon as
\begin{eqnarray}
\phi(r_+)=0,\ \ \ \ \psi(r_+)=r^2_+\frac{f'(r_+)\psi'(r_+)}{m^2}.\label{DCP-bc}
\end{eqnarray}
This regularity condition suggests $\phi_0=0$ and $\psi_0=r^2_+\frac{f_1\psi_1}{m^2}$.

The spacetime geometry dual to the confinement phase is the GB-AdS soliton solution \cite{Cai-Kim2007} which is obtained via analytically continuing the planar GB-AdS black hole solution \cite{Cai2002} as
\begin{eqnarray}
ds^2=r^2\left(-dt^2+h_{ij}dx^idx^j+f(r)dy^2\right)+\frac{dr^2}{r^2f(r)},\label{AdSs}
\end{eqnarray}
where
\begin{eqnarray}
f(r)=\frac{1}{2\alpha}\left[1-\sqrt{1-4\alpha\left(1-\frac{r^5_0}{r^5}\right)}\right],\ \ \ \ r_0=\frac{2}{5R_y},\label{fr-soliton}
\end{eqnarray}
with $r=r_0$ to be a conical singularity of the GB-AdS soliton solution which is removed by imposing a suitable period condition for the coordinate $y$. In this configuration of the spacetime geometry, the equations of motion for $\phi(r)$ and $\psi(r)$ are obtained as
\begin{eqnarray}
\phi''(r)+\left[\frac{f'(r)}{f(r)}+\frac{4}{r}\right]\phi'(r)-\frac{2q^2\psi^2(r)}{r^2f(r)}\phi(r)&=&0,\label{Sol-phi-Eq}\\
\psi''(r)+\left[\frac{f'(r)}{f(r)}+\frac{6}{r}\right]\psi'(r)+\frac{1}{r^2f(r)}\left[\frac{q^2\phi^2(r)}{r^2}-m^2\right]\psi(r)&=&0.\label{Sol-psi-Eq}
\end{eqnarray}
The asymptotic behavior of the matter fields near the AdS boundary is the same as Eq. (\ref{phi-psi-asy-beh}). Whereas, the solution near the tip $r=r_0$ has the following expansions
\begin{eqnarray}
\phi(r)&=&\phi_0+\phi_1\log(r-r_0)+\phi_2(r-r_0)+\phi_3(r-r_0)^2+\cdots,\nonumber\\
\psi(r)&=&\psi_0+\psi_1\log(r-r_0)+\psi_2(r-r_0)+\psi_3(r-r_0)^2+\cdots.\label{MFexpanSol}
\end{eqnarray}
We impose the Neumann-like boundary condition on the matter fields, i.e. $\phi_1=0$ and $\psi_1=0$, to ensure that their value is finite at the tip $r=r_0$. The boundary condition at the tip $r=r_0$ for the matter fields is
\begin{eqnarray}
\phi'(r_0)&=&\frac{2q^2\psi^2(r_0)}{r^2_0f'(r_0)}\phi(r_0),\nonumber\\
\psi'(r_0)&=&-\frac{1}{r^2_0f'(r_0)}\left[\frac{q^2\phi^2(r_0)}{r^2_0}-m^2\right]\psi(r_0).
\end{eqnarray}
By using the expression of $f(r)$ and the expansions of $\phi(r)$ and $\psi(r)$, given in Eqs. (\ref{fr-soliton}) and (\ref{MFexpanSol}), respectively, the above boundary condition becomes
\begin{eqnarray}
\phi_2&=&\frac{2q^2\psi^2_0}{5r_0}\phi_0,\nonumber\\
\psi_2&=&-\frac{1}{5r_0}\left(\frac{q^2\phi^2_0}{r^2_0}-m^2\right)\psi_0,
\end{eqnarray}
which suggests that Eqs. (\ref{Sol-phi-Eq}) and (\ref{Sol-psi-Eq}) allow the solution with $\phi(r_0)\neq0$.

\section{\label{HCSC} Holographic CSC in EGB gravity}
As we discussed above, in the limit that the chemical potential approaches the critical value $\mu_c$, the backreaction of the scalar field is negligible. Since the bulk background configuration is determined by the following action
\begin{equation}
S'_{\text{bulk}}=\int
d^6x\sqrt{-g}\left[R-2\Lambda+\widetilde{\alpha}\mathcal{L}_{GB}-\frac{1}{4}F_{\mu\nu}F^{\mu\nu}\right].\label{EGB-ED-adS-neg}
\end{equation}
The spacetime metric solution dual to the confinement phase is given by the GB-AdS soliton mentioned in the previous section with the constant potential of the gauge field as
\begin{eqnarray}
\phi(r)=\mu.\label{VPSo}
\end{eqnarray}
Whereas, the spacetime metric solution dual to the deconfinement phase is given by the planar GB-RN-AdS black hole with the line element described by Eqs. (\ref{BHa}) and (\ref{GBRNAdSfr}), and the corresponding potential of the gauge field is
\begin{eqnarray}
\phi(r)=\mu\left(1-\frac{r^3_+}{r^3}\right).\label{VFpot}
\end{eqnarray}
The Hawking temperature of the planar GB-RN-AdS black hole is given by
\begin{eqnarray}
T=\frac{1}{4\pi}\left(5r_+-\frac{9\mu^2}{8r_+}\right).
\end{eqnarray}
The non-negative condition of the temperature suggests the suitable region for $\mu/r_+$ as
\begin{eqnarray}
0\leq\frac{\mu}{r_+}\leq\frac{\sqrt{40}}{3}.
\end{eqnarray}

\begin{figure}[t]
 \centering
\begin{tabular}{cc}
\includegraphics[width=0.5 \textwidth]{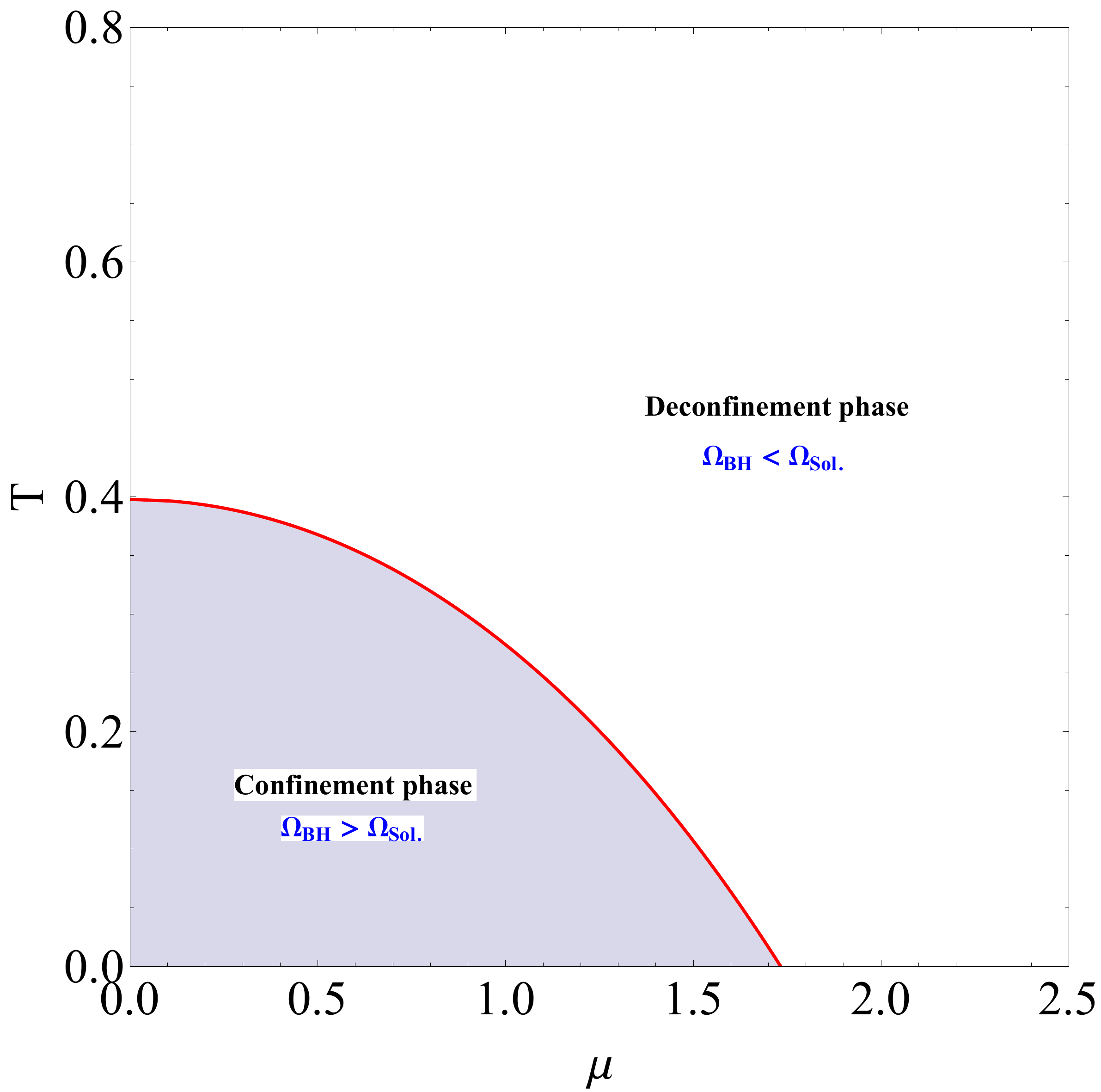}
\end{tabular}
 \caption{The phase diagram for the confinement and deconfinement phases.}\label{C-DC-phases}
\end{figure}

Let us study the phase structure of the bulk background configuration by examining the free energy (in the canonical ensemble) of the planar GB-RN-AdS black hole and GB-AdS soliton. Using the result in Ref. \cite{Miskovic2011}, we can find the total on-shell Euclidean action for the EGB gravity coupled to the $U(1)$ gauge field in the present work as
\begin{eqnarray}
S_E=\left[\left(r^2f\right)'\left(r^4-4\alpha r^4f\right)\Big|^\infty_{r_+}-l^4_{\text{eff}}\left(1-\frac{4\alpha}{l^2_{\text{eff}}}\right)r^4f^2\left(r^2f\right)'\Big|^\infty-r^4\phi\phi'\Big|^\infty_{r_+}\right]\frac{4\pi}{5r_0}\frac{V_3}{T},\label{OSEA}
\end{eqnarray}
where $V_3=\int dx_1dx_2dx_3$. Then, we obtain the free energy of the planar GB-RN-AdS black hole and GB-AdS soliton as
\begin{eqnarray}
\Omega_{\text{BH}}&=&-r^5_+\left(1+\frac{3\mu^2}{8r^2_+}\right)\frac{4\pi}{5r_0}V_3,\\
\Omega_{\text{Sol.}}&=&-r^5_0\frac{4\pi}{5r_0}V_3.
\end{eqnarray}
Here, we see that the free energies of the planar GB-RN-AdS black hole and GB-AdS soliton in Einstein gravity and the EGB gravity are the same in the planar case although the solutions in these two kinds of gravity are different. By comparing their free energy, one finds which configuration is thermodynamically favored. The corresponding phase diagram is depicted in Fig. \ref{C-DC-phases}. 
The critical curve (red one) which separates the configuration of the GB-AdS soliton and that of the planar GB-RN-AdS black hole is determined by the equation $\Omega_{\text{BH}}=\Omega_{\text{Sol.}}$.  

In the following, we study how the phase structure of the bulk background configuration, mentioned above, changes when the scalar field condensate appears.

\subsection{Deconfinement phase}

\begin{figure}[t]
 \centering
\begin{tabular}{cc}
\includegraphics[width=0.45 \textwidth]{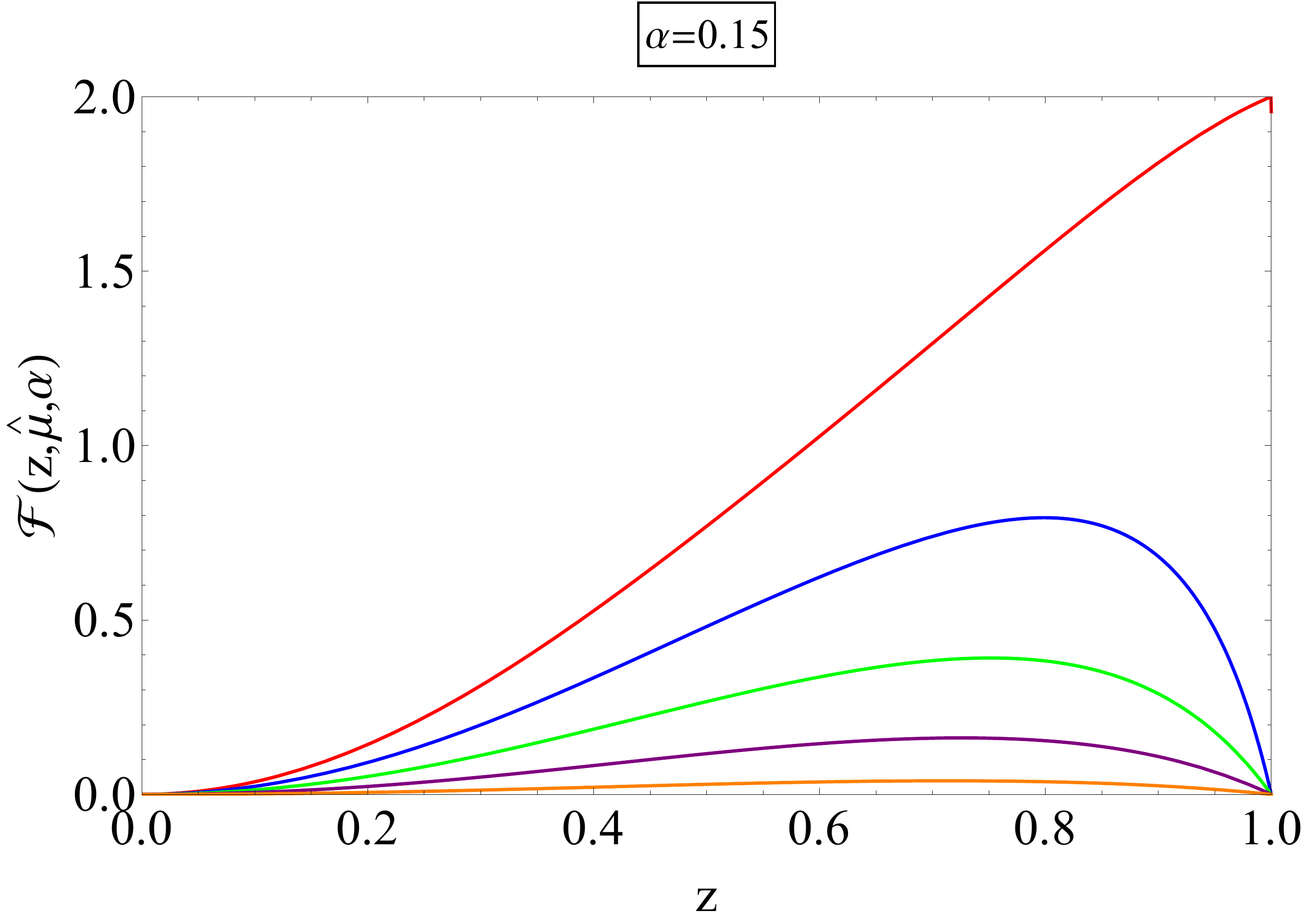}
\hspace*{0.05\textwidth}
\includegraphics[width=0.45 \textwidth]{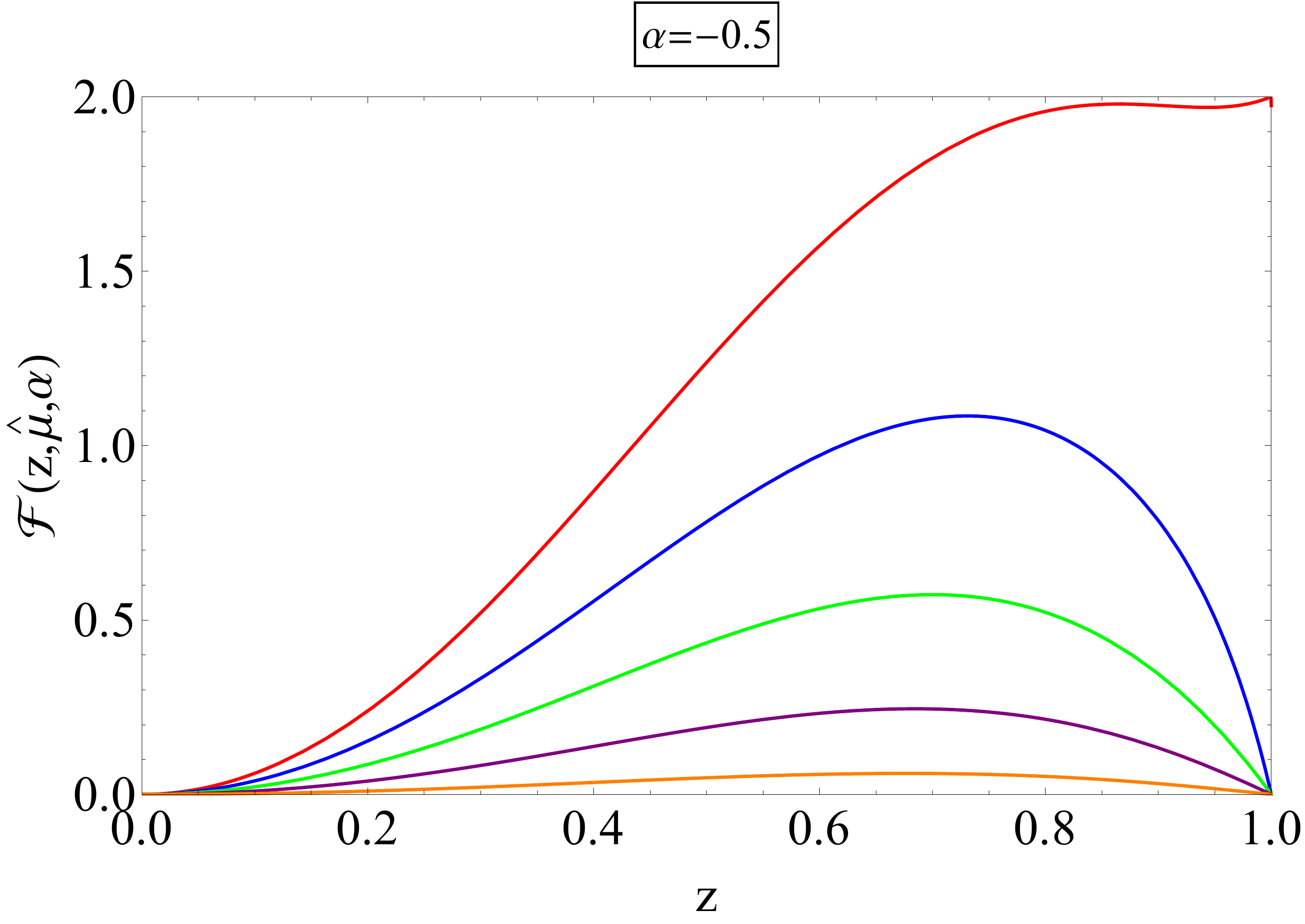}\\
\includegraphics[width=0.45 \textwidth]{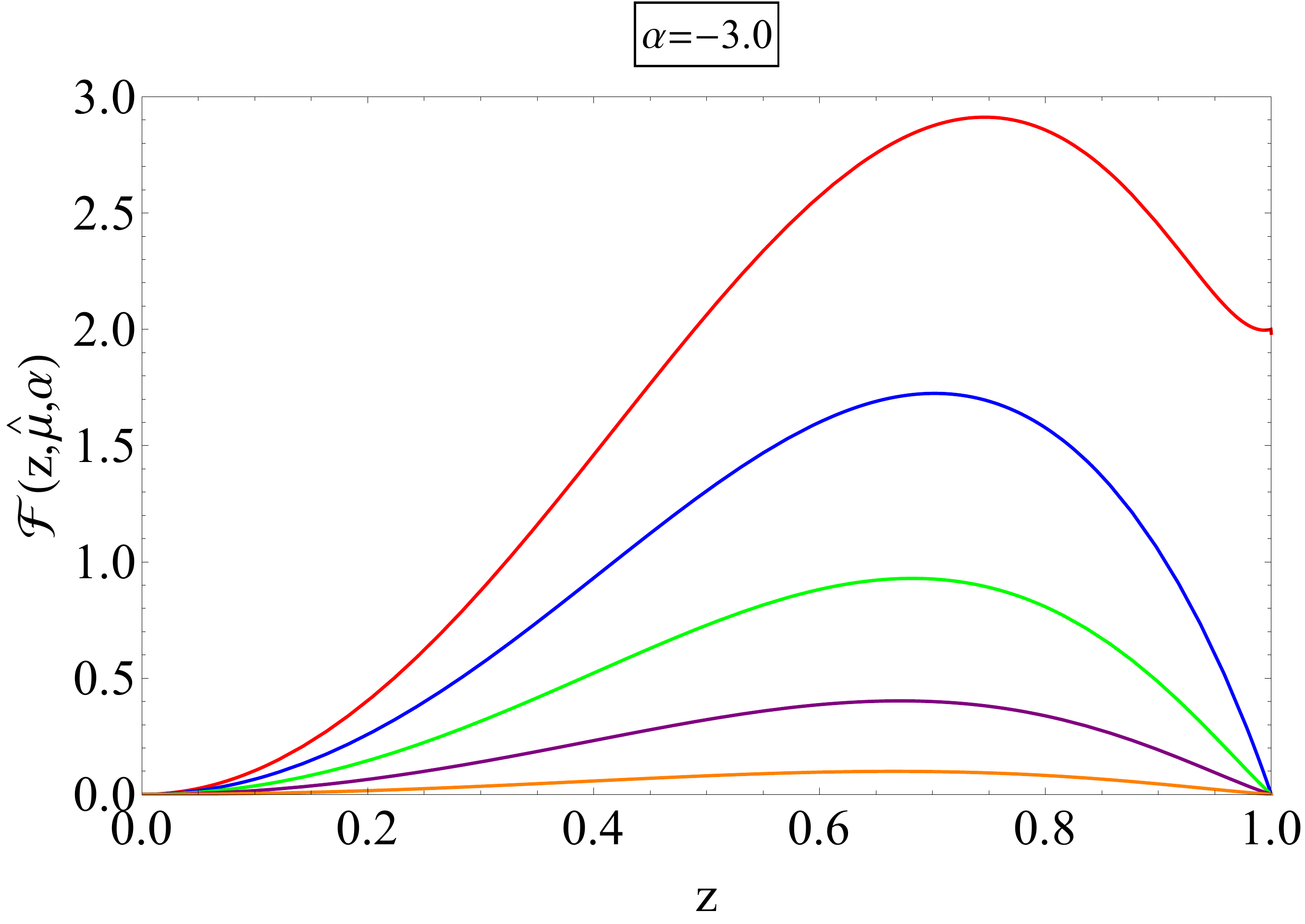}
\hspace*{0.05\textwidth}
\includegraphics[width=0.45 \textwidth]{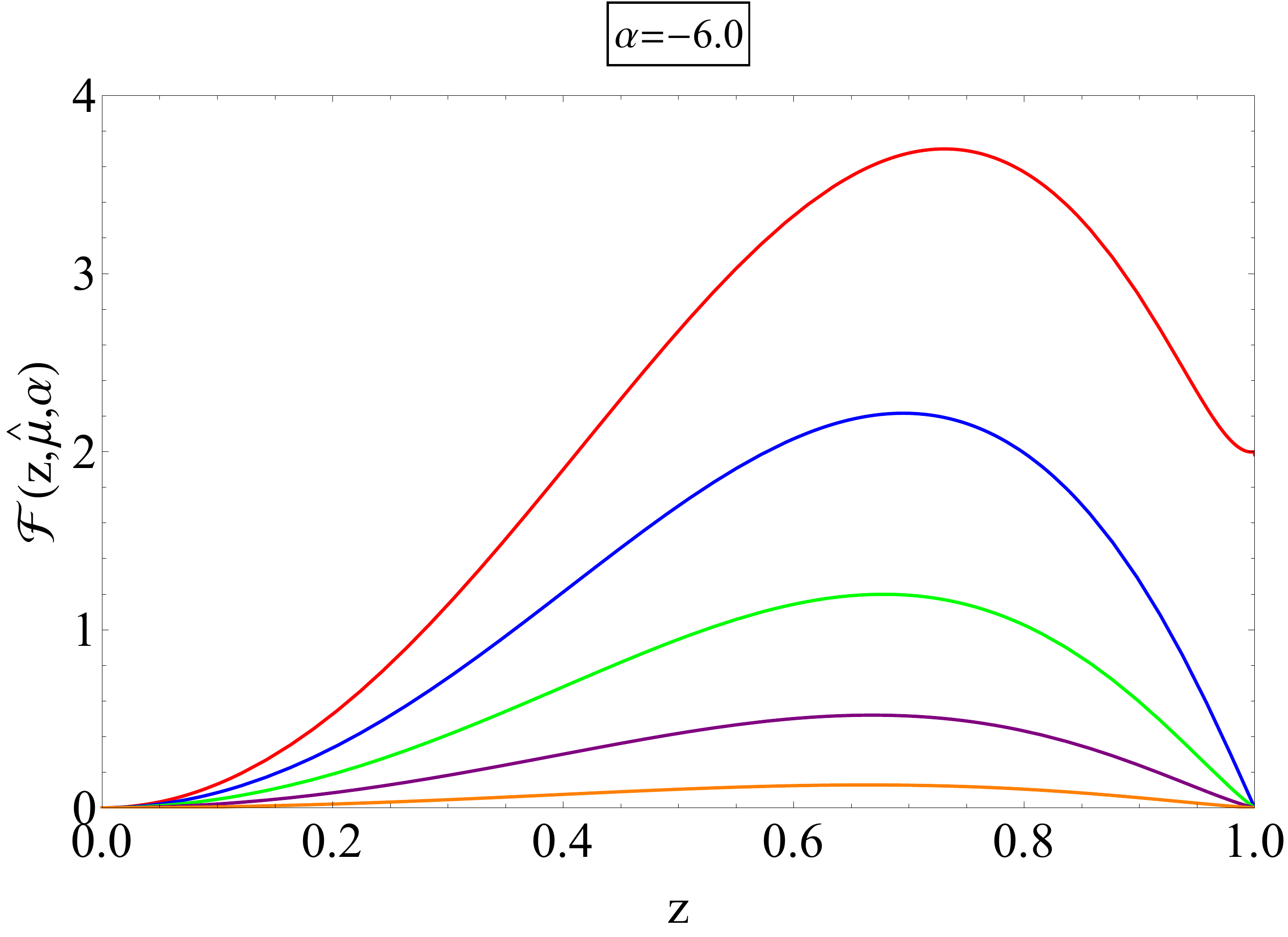}
\end{tabular}
  \caption{The behavior of $\mathcal{F}(z,\hat{\mu},\alpha)$ as a function of $z$ with various values of $\hat{\mu}$ and $\alpha$. The red, blue, green, purple, and orange curves correspond to $\hat{\mu}=\frac{\sqrt{40}}{3}$, $0.8\times\frac{\sqrt{40}}{3}$, $0.6\times\frac{\sqrt{40}}{3}$, $0.4\times\frac{\sqrt{40}}{3}$, $0.2\times\frac{\sqrt{40}}{3}$, respectively.}\label{FfunB-A}
\end{figure}

\begin{figure}[t]
 \centering
\begin{tabular}{cc}
\includegraphics[width=0.45 \textwidth]{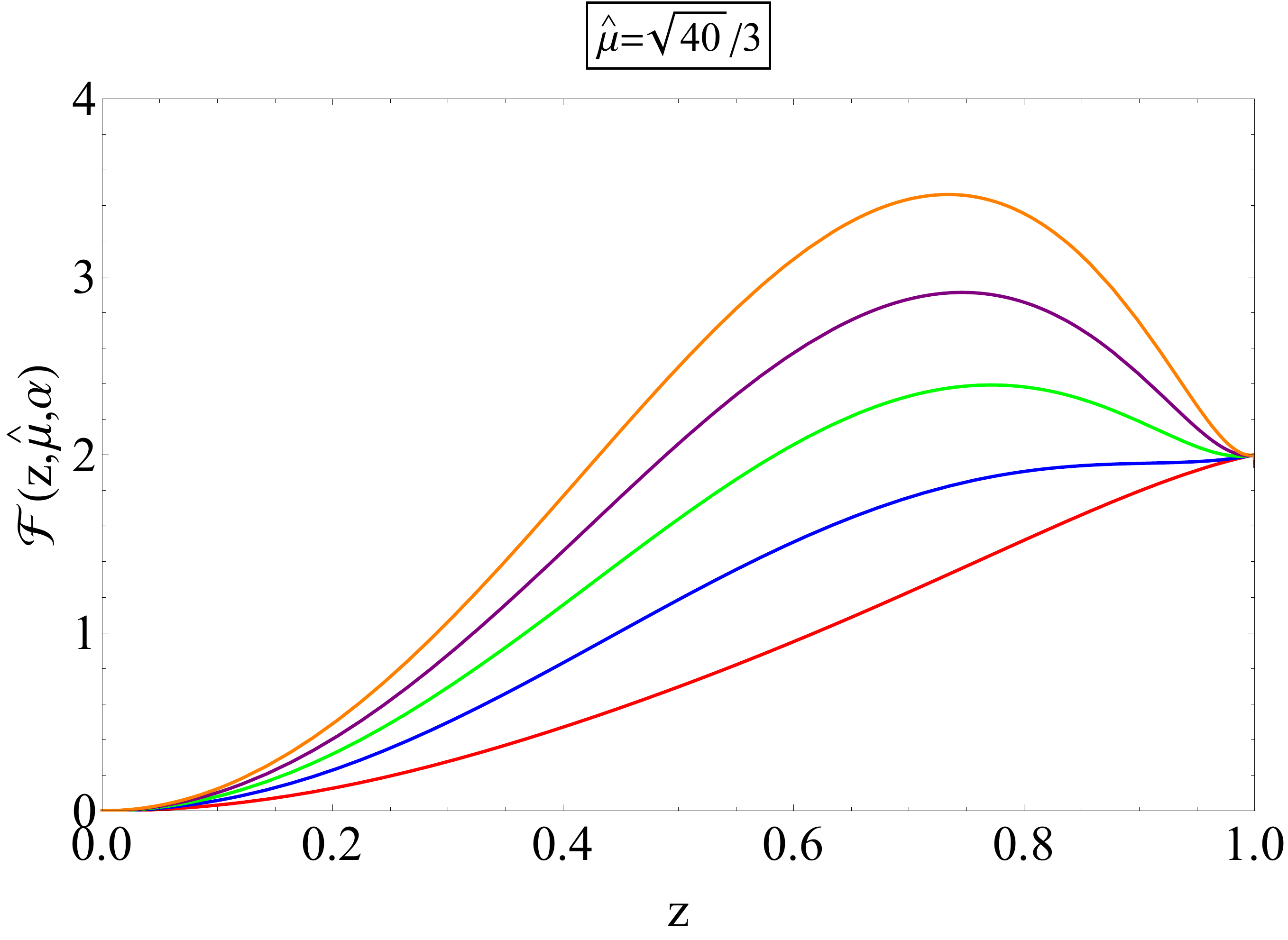}
\hspace*{0.05\textwidth}
\includegraphics[width=0.45 \textwidth]{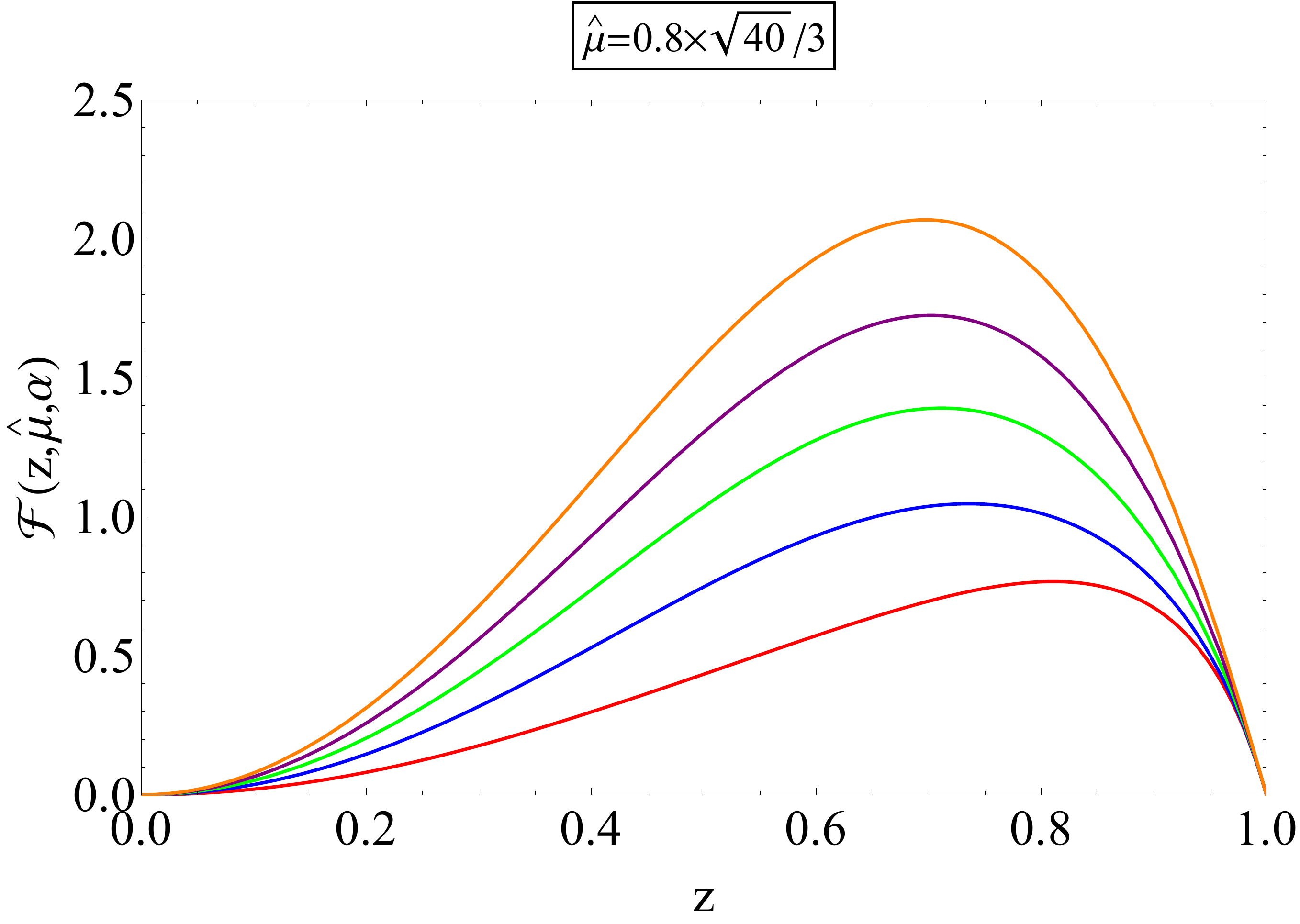}\\
\includegraphics[width=0.45 \textwidth]{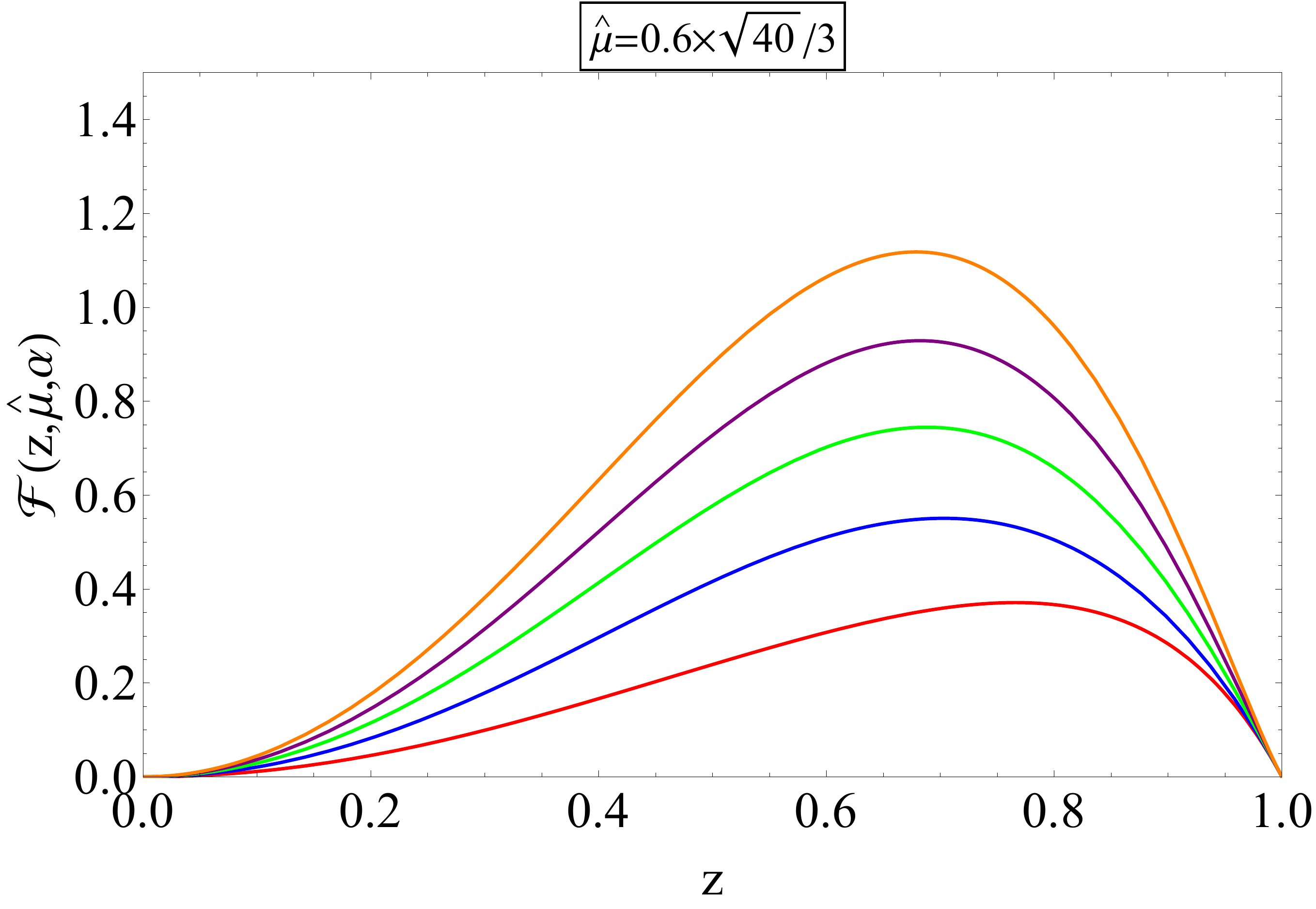}
\hspace*{0.05\textwidth}
\includegraphics[width=0.45 \textwidth]{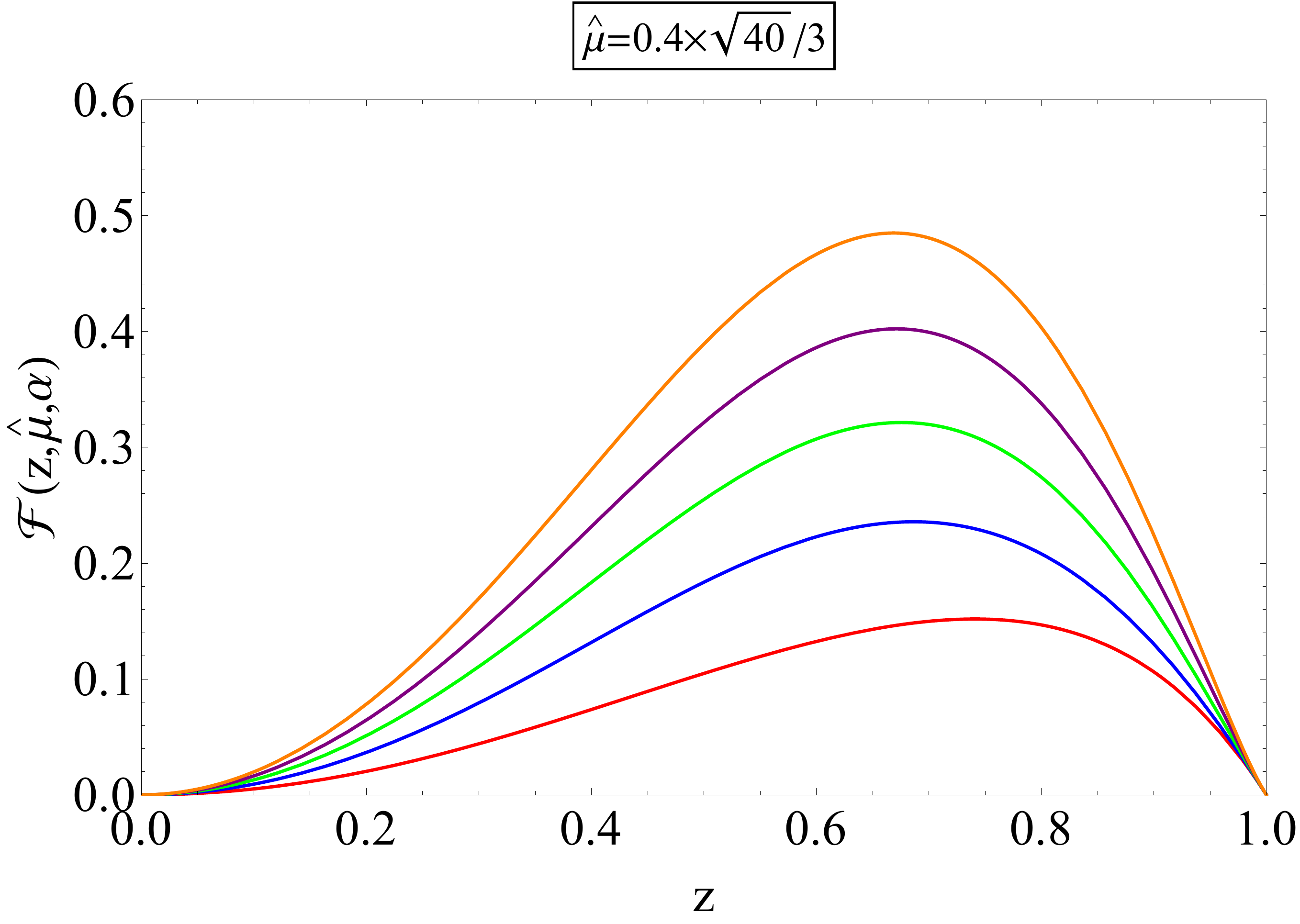}
\end{tabular}
  \caption{The behavior of $\mathcal{F}(z,\hat{\mu},\alpha)$ as a function of $z$ with various values of $\hat{\mu}$ and $\alpha$. The red, blue, green, purple, and orange curves correspond to $\alpha=0.2$, $-0.4$, $-1.5$, $-3.0$, $-5.0$, respectively.}\label{FfunB-B}
\end{figure}

\begin{figure}[t]
 \centering
\begin{tabular}{cc}
\includegraphics[width=0.45 \textwidth]{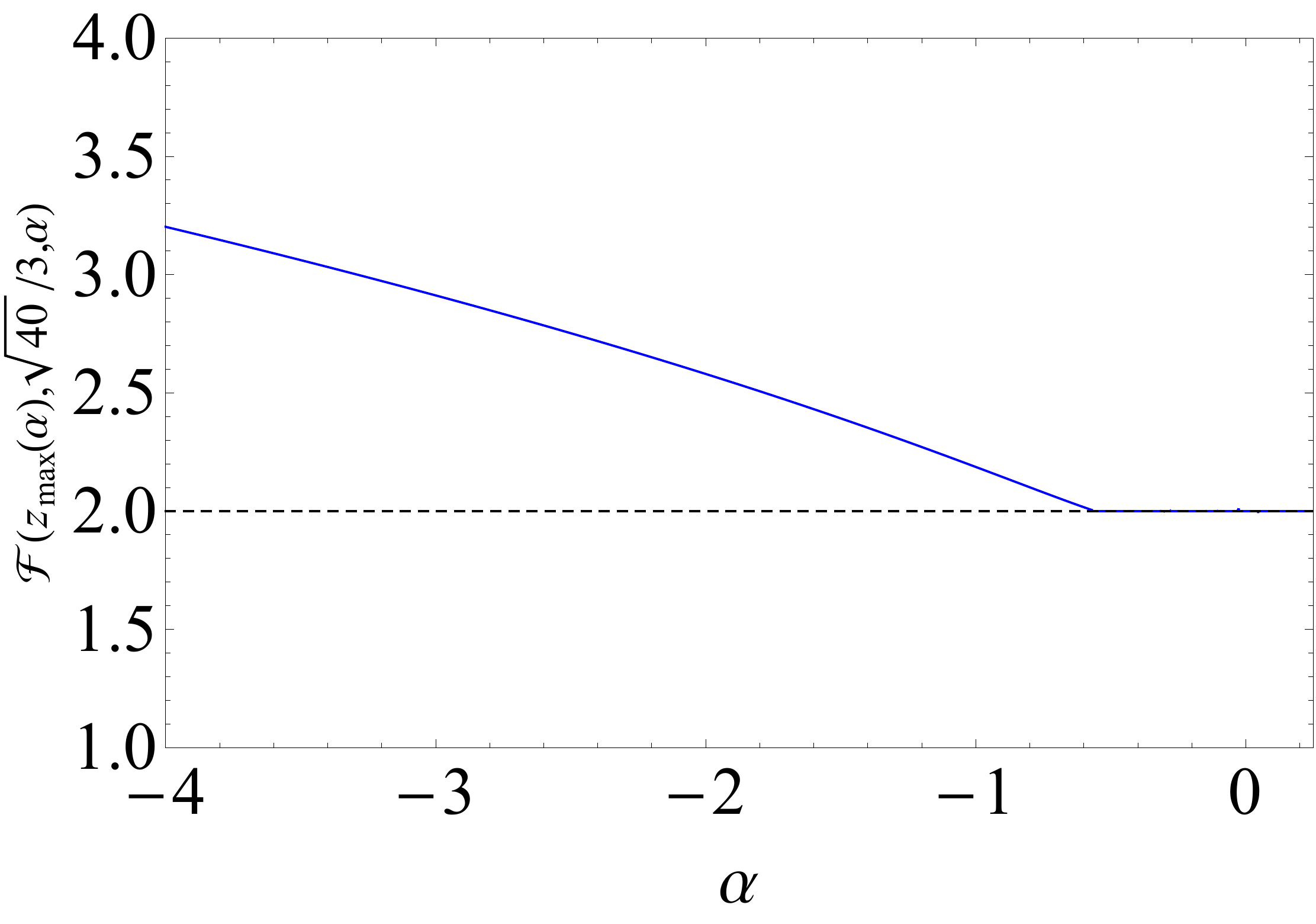}
\hspace*{0.05\textwidth}
\includegraphics[width=0.45 \textwidth]{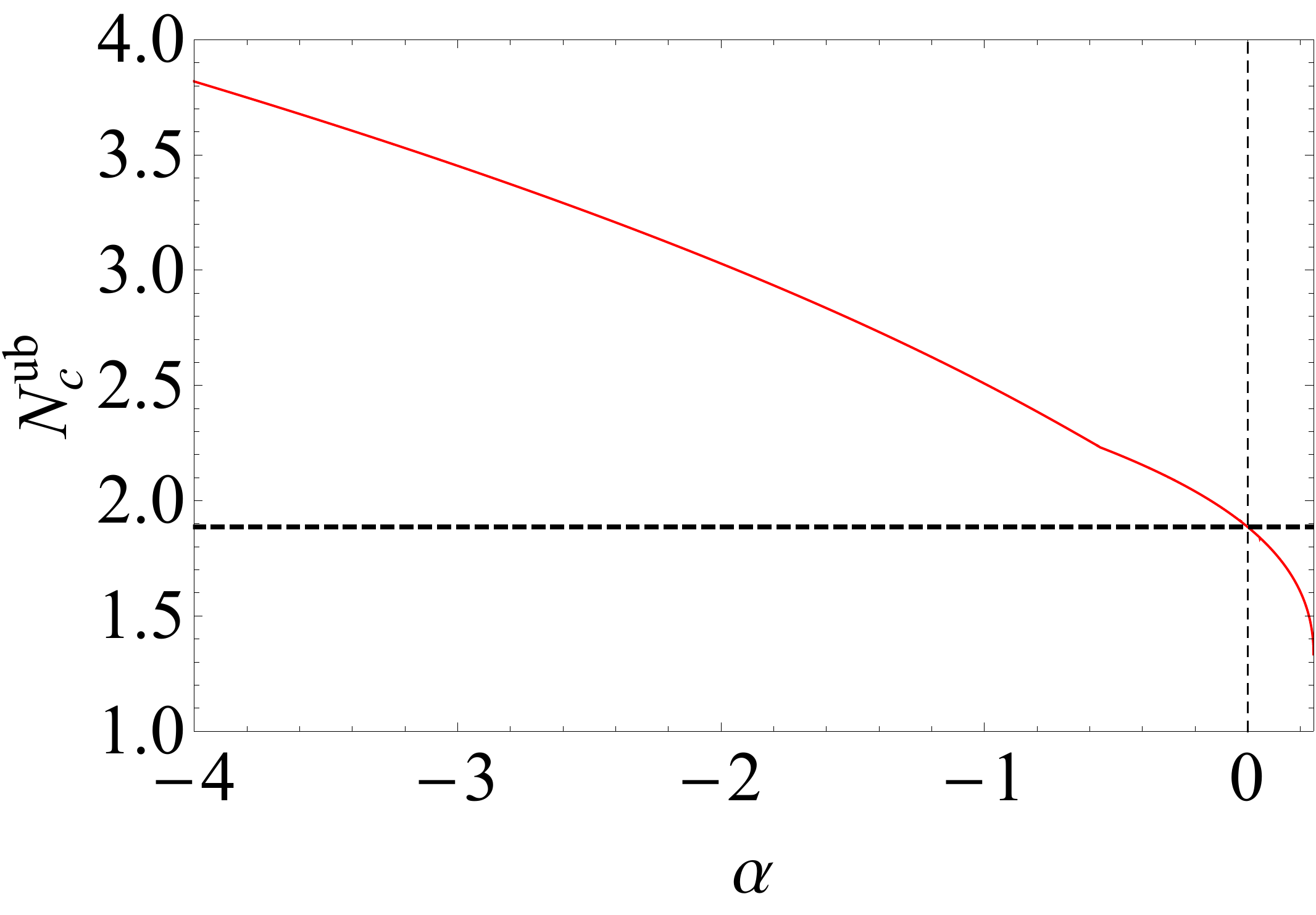}\\
\end{tabular}
  \caption{Left panel: The maximal value $\mathcal{F}(z_{\text{max}}(\alpha),\sqrt{40}/3,\alpha)$ as a function of $\alpha$. Right panel: The upper bound for $N_c$ as a function of $\alpha$. The horizontal dashed black lines correspond to the case of Einstein gravity.}\label{Fmax-Nc}
\end{figure}

Let us first study the necessary condition which destabilizes the scalar field and makes the condensation occurring. From the equation of motion for the scalar field, one can find the effective squared mass $m^2_{\text{eff}}$ of the scalar field as
\begin{eqnarray}
m^2_{\text{eff}}=m^2-\frac{q^2\phi^2(r)}{r^2f(r)},
\end{eqnarray}
with $f(r)$ and $\phi(r) $ given by Eqs. (\ref{GBRNAdSfr}) and (\ref{VFpot}), respectively. The necessary condition which $m^2_{\text{eff}}$ breaks the BF bound is given as
\begin{eqnarray}
m^2_{\text{eff}}<-\frac{25}{4l^2_{\text{eff}}},
\end{eqnarray}
which leads to
\begin{eqnarray}
\frac{q^2\phi^2(r)}{r^2f(r)}>\frac{9}{4l^2_{\text{eff}}}.\label{deleq}
\end{eqnarray}
The left-hand side of (\ref{deleq}) can be rewritten as
\begin{eqnarray}
\frac{q^2\phi^2(r)}{r^2f(r)}=q^2\frac{2\alpha z^2(1-z^3)^2\hat{\mu}^2}{1-\sqrt{1-4\alpha(1-z^5)+\frac{3\alpha\hat{\mu}^2}{2}z^5(1-z^3)}}\equiv q^2\mathcal{F}(z,\hat{\mu},\alpha),
\end{eqnarray}
where $z\equiv r_+/r$ and $\hat{\mu}\equiv\mu/r_+$. The behavior of the function $\mathcal{F}(z,\hat{\mu},\alpha)$ in terms of $z$, $\hat{\mu}$, and $\alpha$ is shown in Figs. \ref{FfunB-A} and \ref{FfunB-B}. 
We can see that $\mathcal{F}(z,\hat{\mu},\alpha)$ increases with either the growth of $\hat{\mu}$ or the decreasing of $\alpha$ at an arbitrary value of $z$. In particular, one finds that $\mathcal{F}(z,\hat{\mu},\alpha)$ would get the maximal value at $\hat{\mu}=\frac{\sqrt{40}}{3}$. In the region of $\alpha\gtrsim-0.56$, the maximal value of $\mathcal{F}(z,\sqrt{40}/3,\alpha)$ is about two and thus there is an upper bound as $\mathcal{F}(z,\hat{\mu},\alpha)<2$. In the remaining region of $\alpha$, $\mathcal{F}(z,\sqrt{40}/3,\alpha)$ gets the maximal value at a point $z_{\text{max}}$ which depends on $\alpha$ and is found by solving the following equation
\begin{eqnarray}
\frac{\partial}{\partial z}\mathcal{F}(z,\sqrt{40}/3,\alpha)=0,
\end{eqnarray}
which leads to
\begin{eqnarray}
3-12\alpha+4z^3\left[(12-2z^2-17z^5+10z^8)\alpha-3\right]+(4z^3-1)\sqrt{9-12(3-8z^5+5z^8)\alpha}=0.
\end{eqnarray}
The corresponding maximal value $\mathcal{F}(z_{\text{max}}(\alpha),\sqrt{40}/3,\alpha)$ as a function of $\alpha$ is numerically given in the left panel of Fig. \ref{Fmax-Nc}. As a result, we obtain the following equality
\begin{eqnarray}
0<\frac{q^2\phi^2(r)}{r^2f(r)}<\frac{4}{N^2_c}\mathcal{F}(z_{\text{max}}(\alpha),\sqrt{40}/3,\alpha),
\end{eqnarray}
where we have used the relation $q=\frac{2}{N_c}$. This result along with (\ref{deleq}) leads to
\begin{eqnarray}
N_c<\frac{4}{3}\sqrt{\frac{2\alpha\mathcal{F}(z_{\text{max}}(\alpha),\sqrt{40}/3,\alpha)}{1-\sqrt{1-4\alpha}}}\equiv N^{{\text{ub}}}_c(\alpha).
\end{eqnarray}
More explicitly, we show the behavior of the upper bound for $N_c$ as a function of the GB coupling parameter $\alpha$ in the right panel of Fig. \ref{Fmax-Nc}. 
We observe that $N^{{\text{ub}}}_c$ decreases with the growth of $\alpha$. In the case of $\alpha=0$ corresponding to Einstein gravity, we obtain the upper bound for the color number $N_c$ as $N_c<\frac{4\sqrt{2}}{3}\simeq1.89$ \cite{Ghoroku2019}. In addition, $N^{{\text{ub}}}_c$ in the EGB gravity with $\alpha>0$ is lower than that in Einstein gravity. This suggests that the scalar field condensate can not be found in the EGB gravity with $\alpha>0$ for $N_c\geq2$. However, in the EGB gravity with $\alpha<0$ the upper bound $N^{{\text{ub}}}_c$ is enhanced compared to Einstein gravity and increases as the magnitude of the GB coupling parameter $\alpha$ grows. As a result, the presence of the GB term can lead the scalar field condensate with $N_c\geq2$ which is realistic to realize the CSC phase in the YM theory.

Above the critical chemical potential $\mu_c$, the CSC phase occurs due to the condensation of the scalar field which corresponds to the nontrivial solution of $\psi$ as $J_C=0$ [to guarantee the spontaneous breaking of the $U(1)$ symmetry in the system] and $C\neq0$. We can obtain the critical chemical potential $\mu_c$ and thus the critical curve in the $\mu-T$ plane by solving numerically Eqs. (\ref{r-f-Eq})$-$(\ref{r-psi-Eq}) using the shooting method. In this method, the boundary values of $\phi$ and $\psi$ can be derived by setting their appropriate value near the event horizon. Of course, the critical chemical potential $\mu_c$ and the critical curve depend on both the GB coupling parameter $\alpha$ and the color number $N_c$.

\begin{figure}[t]
 \centering
\begin{tabular}{cc}
\includegraphics[width=0.45 \textwidth]{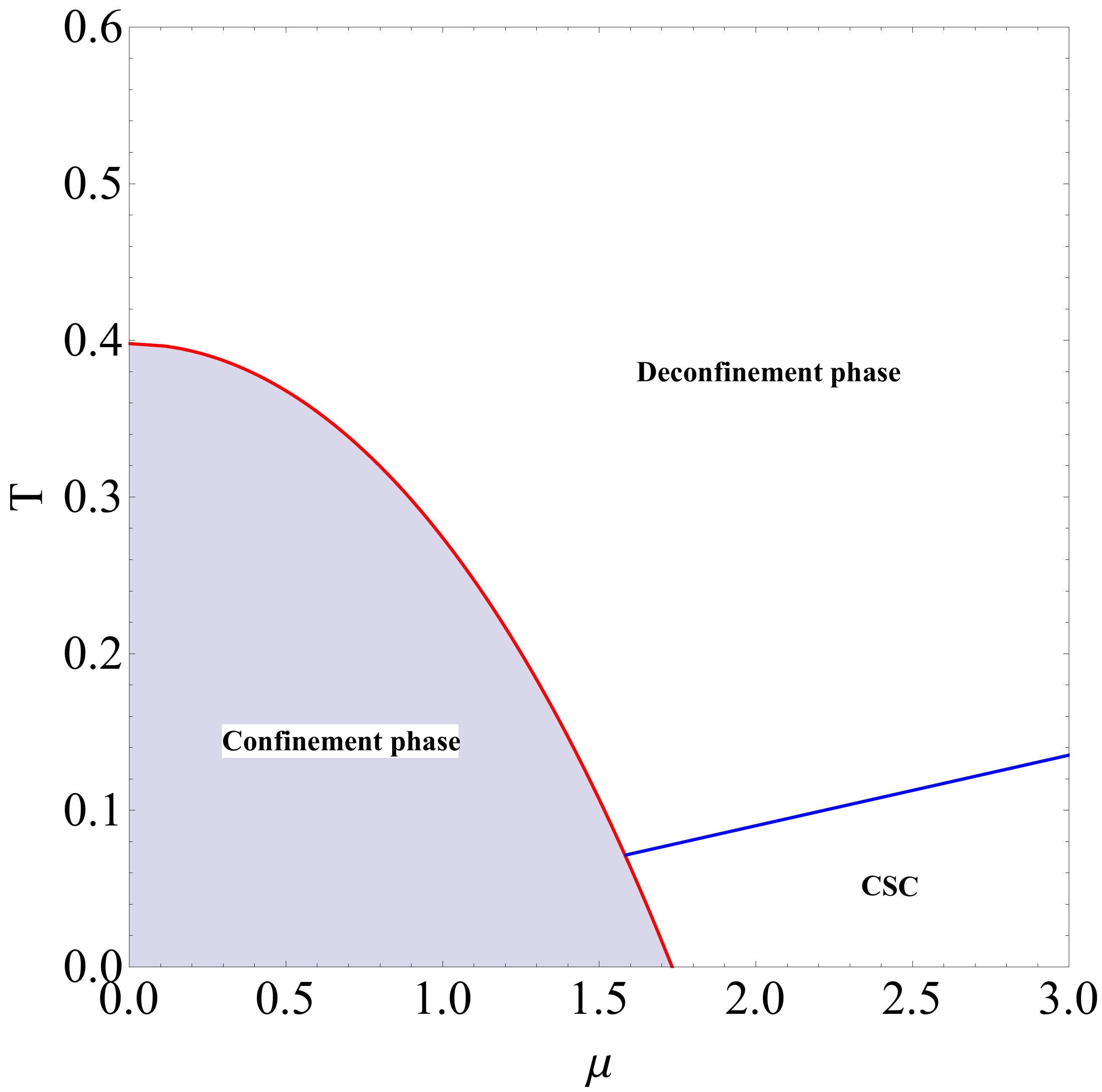}
\hspace*{0.05\textwidth}
\includegraphics[width=0.45 \textwidth]{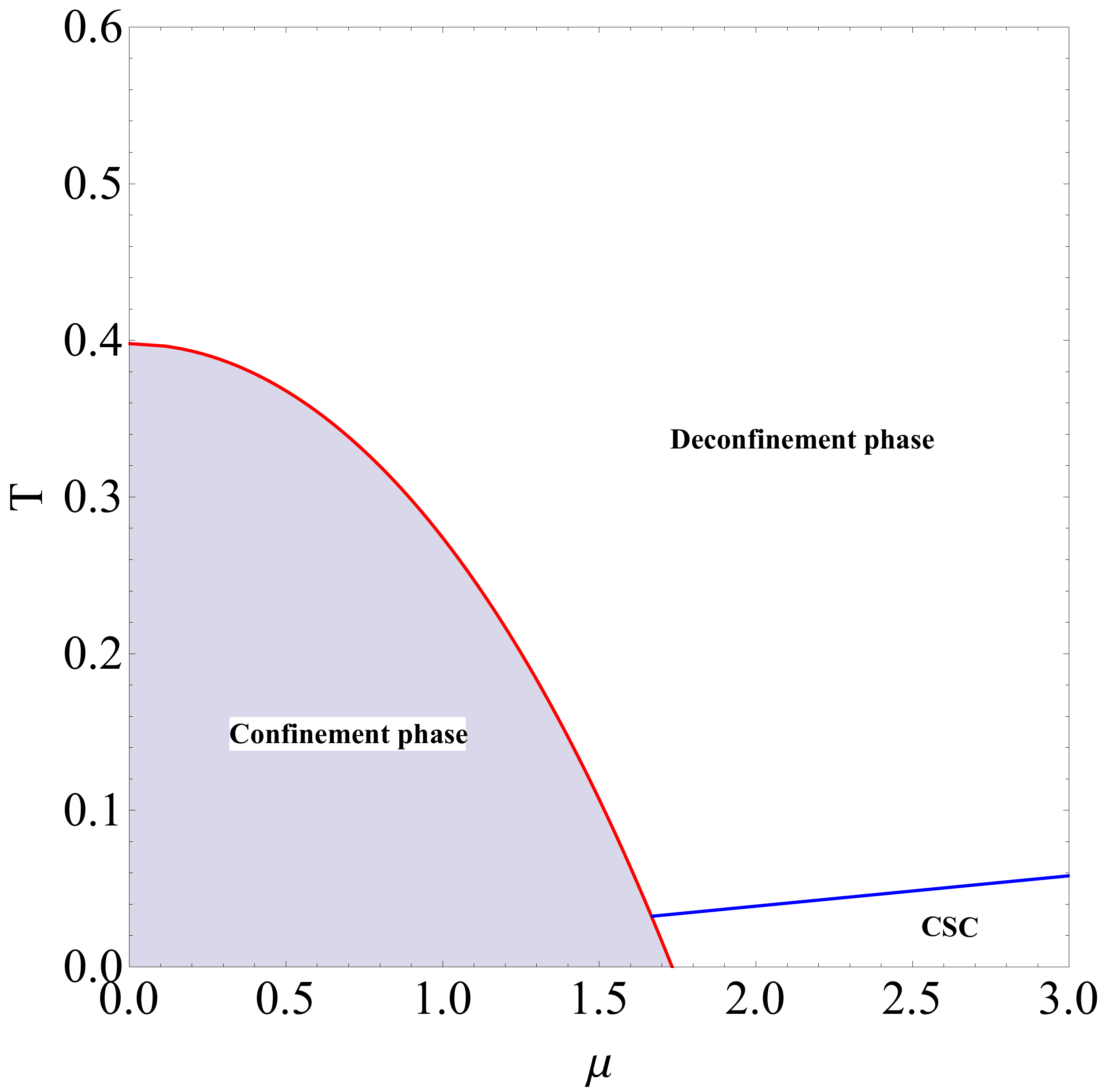}\\
\includegraphics[width=0.45 \textwidth]{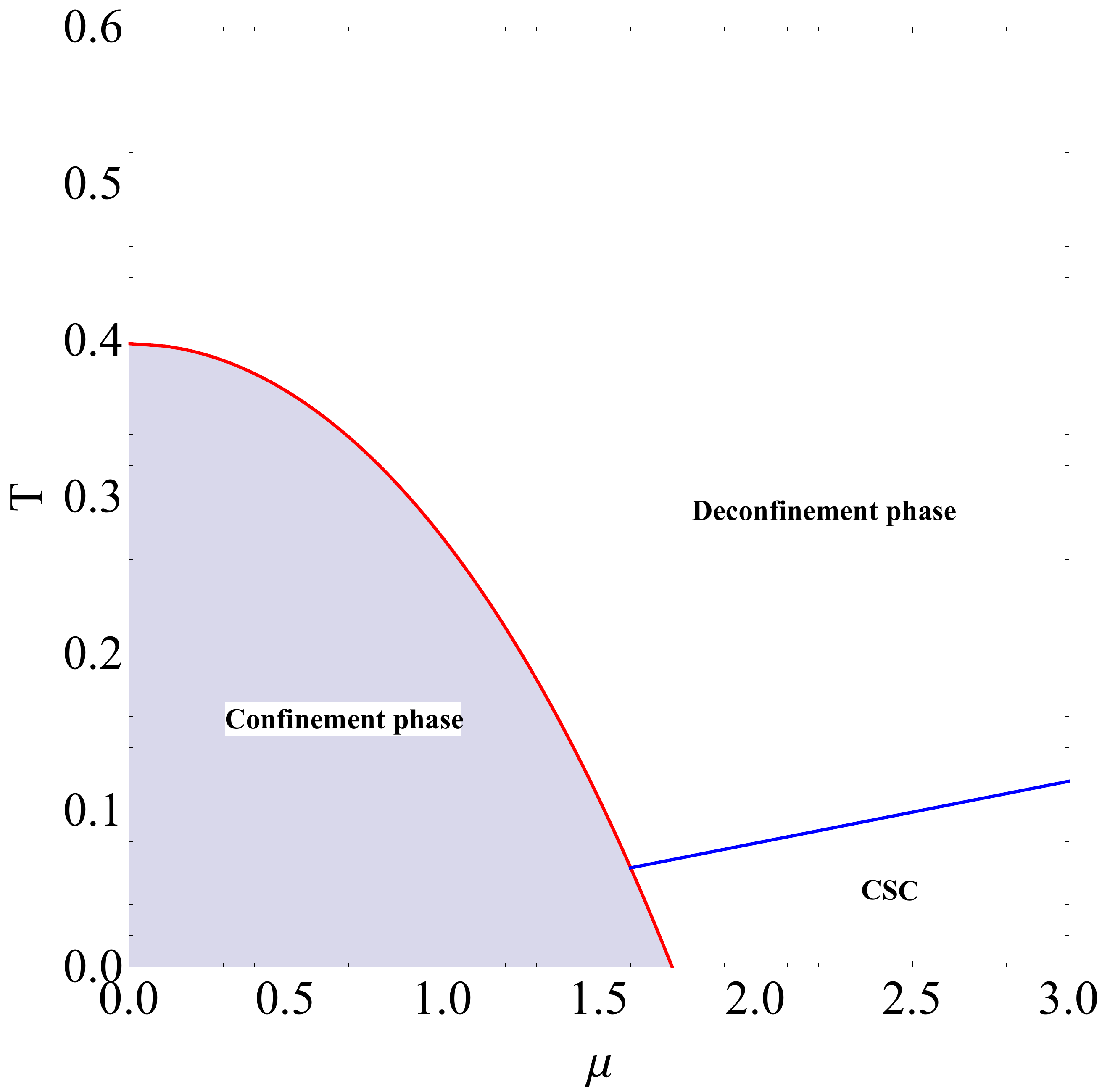}
\hspace*{0.05\textwidth}
\includegraphics[width=0.45 \textwidth]{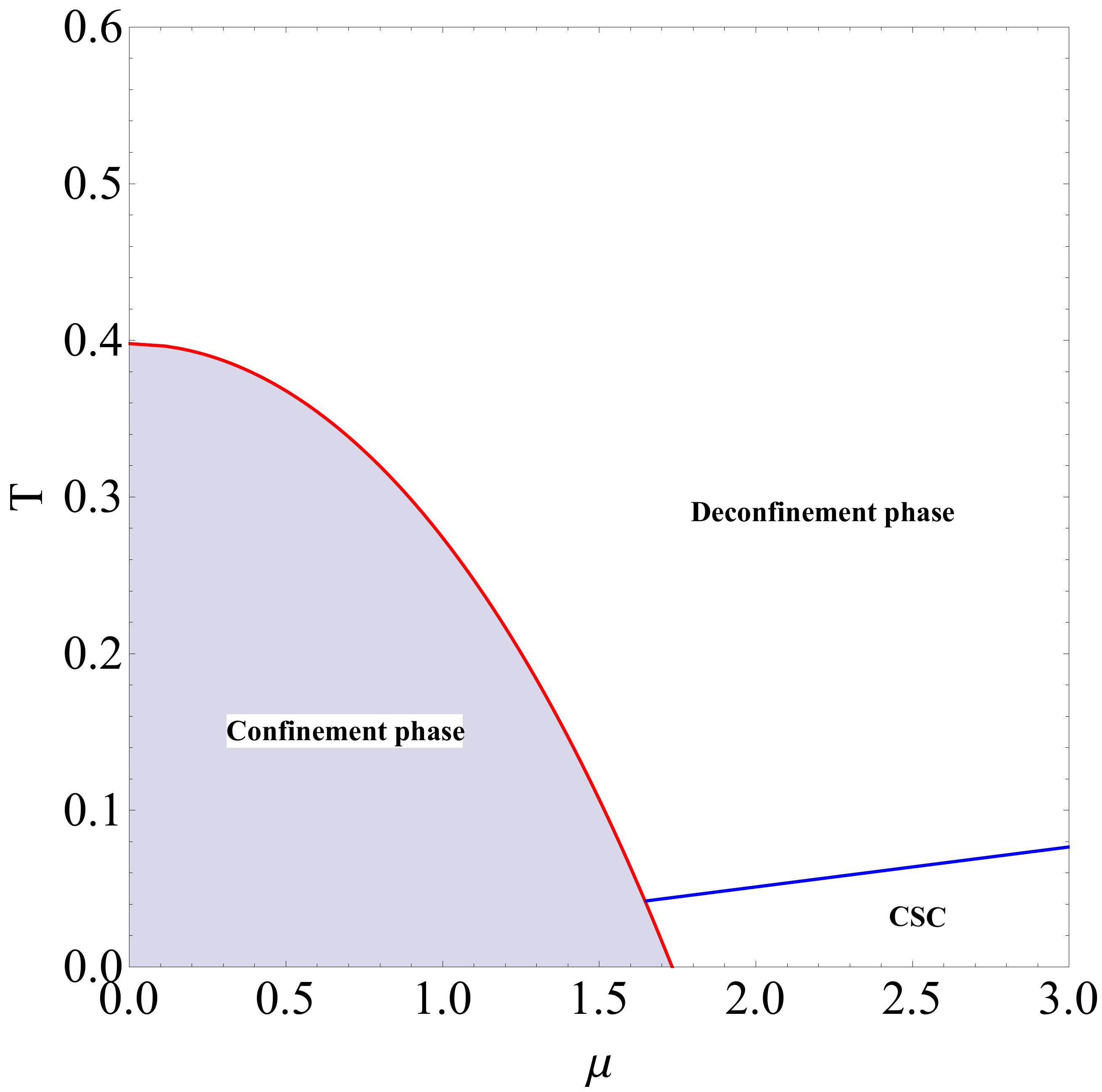}
\end{tabular}
  \caption{The phase diagram in the case of that the scalar field is taken into account for various values of $N_c$ and $\alpha$. Top-left panel: $N_c=2$ and $\alpha=-3.2$. Top-right panel: $N_c=2$ and $\alpha=-2.4$. Bottom-left panel: $N_c=3$ and $\alpha=-9.0$. Bottom-right panel: $N_c=3$ and $\alpha=-8.0$. The regions below the blue lines refer to the CSC phase.}\label{PD-Nc23}
\end{figure}

As analyzed above, the EGB gravity with $\alpha<0$ can lead to the scalar field condensate with $N_c\geq2$. Thus, we solve numerically Eqs. (\ref{r-f-Eq})$-$(\ref{r-psi-Eq}) with the negative GB coupling parameter to find the critical chemical potential for $N_c\geq2$ and thus the corresponding phase diagram. We show the numerical values for the scaled critical chemical potential $\mu_c/r_+$ and the slope of the critical line $T_c=T_c(\mu_c)$ for various values of the GB coupling parameter $\alpha$ for $N_c=2$ and $N_c=3$ in Tables \ref{Nc2} and \ref{Nc3}, respectively.
\begin{table}[!htp]
\centering
\begin{tabular}{c|c|c}
  \hline
  \hline
  $\alpha$ & $\mu_c/r_+$ & $T_c/\mu_c$ \\
  \hline
  $\ \ \ \ \ \ -2.0 \ \ \ \ \ \ $ & $\ \ \ \ \ \ 2.06670 \ \ \ \ \ \ $ & $\ \ \ \ \ \ 0.00750 \ \ \ \ \ \ $\\
  \hline
  $\ \ \ \ \ \ -2.4 \ \ \ \ \ \ $ & $\ \ \ \ \ \ 2.00273 \ \ \ \ \ \ $ & $\ \ \ \ \ \ 0.01938 \ \ \ \ \ \ $\\
  \hline
  $\ \ \ \ \ \ -2.8 \ \ \ \ \ \ $ & $\ \ \ \ \ \ 1.93572 \ \ \ \ \ \ $ & $\ \ \ \ \ \ 0.03225 \ \ \ \ \ \ $\\
  \hline
  $\ \ \ \ \ \ -3.2 \ \ \ \ \ \ $ & $\ \ \ \ \ \ 1.87150 \ \ \ \ \ \ $ & $\ \ \ \ \ \ 0.04506 \ \ \ \ \ \ $\\
  \hline
  $\ \ \ \ \ \ -3.8 \ \ \ \ \ \ $ & $\ \ \ \ \ \ 1.78346 \ \ \ \ \ \ $ & $\ \ \ \ \ \ 0.06343 \ \ \ \ \ \ $\\
  \hline
  $\ \ \ \ \ \ -4.2 \ \ \ \ \ \ $ & $\ \ \ \ \ \ 1.73046 \ \ \ \ \ \ $ & $\ \ \ \ \ \ 0.07501 \ \ \ \ \ \ $\\
  \hline
  \hline
\end{tabular}
\caption{The numerical values for $\mu_c/r_+$ and $T_c/\mu_c$ with various values of $\alpha$ at $N_c=2$.} \label{Nc2}
\end{table}
\begin{table}[!htp]
\centering
\begin{tabular}{c|c|c}
  \hline
  \hline
  $\alpha$ & $\mu_c/r_+$ & $T_c/\mu_c$ \\
  \hline
  $\ \ \ \ \ \ -6.5 \ \ \ \ \ \ $ & $\ \ \ \ \ \ 2.08324 \ \ \ \ \ \ $ & $\ \ \ \ \ \ 0.00449 \ \ \ \ \ \ $\\
  \hline
  $\ \ \ \ \ \ -7.0 \ \ \ \ \ \ $ & $\ \ \ \ \ \ 2.04613 \ \ \ \ \ \ $ & $\ \ \ \ \ \ 0.01128 \ \ \ \ \ \ $\\
  \hline
  $\ \ \ \ \ \ -7.5 \ \ \ \ \ \ $ & $\ \ \ \ \ \ 2.00801 \ \ \ \ \ \ $ & $\ \ \ \ \ \ 0.01838 \ \ \ \ \ \ $\\
  \hline
  $\ \ \ \ \ \ -8.0 \ \ \ \ \ \ $ & $\ \ \ \ \ \ 1.97044 \ \ \ \ \ \ $ & $\ \ \ \ \ \ 0.02552 \ \ \ \ \ \ $\\
  \hline
  $\ \ \ \ \ \ -8.5 \ \ \ \ \ \ $ & $\ \ \ \ \ \ 1.93404 \ \ \ \ \ \ $ & $\ \ \ \ \ \ 0.03259 \ \ \ \ \ \ $\\
  \hline
  $\ \ \ \ \ \ -9.0 \ \ \ \ \ \ $ & $\ \ \ \ \ \ 1.89902 \ \ \ \ \ \ $ & $\ \ \ \ \ \ 0.03951 \ \ \ \ \ \ $\\
  \hline
  \hline
\end{tabular}
\caption{The numerical values for $\mu_c/r_+$ and $T_c/\mu_c$ with various values of $\alpha$ at $N_c=3$.} \label{Nc3}
\end{table}
It is found that as the magnitude of the GB coupling parameter $\alpha$ increases, the critical chemical potential $\mu_c$ decreases for the event horizon $r_+$ kept fixed. This means that the condensation of the scalar field is easier to form with increasing the magnitude of $\alpha$. In addition, we observe that the larger magnitude of $\alpha$ leads to the larger slope of the critical line $T_c=T_c(\mu_c)$. This suggests that the region of the CSC phase is larger, as seen in Fig. \ref{PD-Nc23} for $N_c=2$ and $N_c=3$. On the other hand, increasing the magnitude of $\alpha$ makes the CSC phase more stable.

In Fig. \ref{PD-Nc23}, we show the phase diagram when takes account of the backreaction of the scalar field for $N_c=2$ (top panels) and $N_c=3$ (bottom panels) with various values of the GB coupling parameter $\alpha$.
The phase diagram is dramatically different from that in the case of the absence of the scalar field, given in Fig. \ref{C-DC-phases}, with the presence of the critical line (the blue lines) in the deconfinement region below which it represents the CSC phase which is dual to the the planar GB-RN-AdS black hole with scalar hair. (As we see later, the CSC state does not exist in the confinement phase with the values of $\alpha$ considered in Fig. \ref{PD-Nc23}.) The free energy of this configuration is given by
\begin{eqnarray}
\Omega^{\text{sh}}_{\text{BH}}&=&\left[-r^5_+\left(1+\frac{3\mu^2}{8r^2_+}\right)-\int^\infty_{r_+}\frac{q^2r^2\phi^2\psi^2}{f(r)}dr\right]\frac{4\pi}{5r_0}V_3.
\end{eqnarray}
The second term in this expression is due to the condensation of the scalar field which is always negative and thus the free energy of the CSC state is always lower than that of the normal deconfinement state. In this way, above the critical chemical potential the CSC state contributes dominantly to the thermodynamics and since it is thermodynamically favored. We conclude that the CSC phase in 4D YM theories with $N_c\geq2$ can exist in the gravitational dual model within the framework of the EGB gravity with $\alpha<0$, which can not be found within the framework of Einstein gravity.

Let us make some comments on the following points. First, we need to understand why the presence of the GB term with the negative GB coupling parameter $\alpha$ can work to realize the CSC phase transition for $N_c\geq2$, which is not found in Einstein gravity. It should be noted that as the color number increases, the charge $q=2/N_c$ of the dual scalar field decreases. This means that, with respect to the holographic model for $N_c\geq2$ in Einstein gravity, the electrostatic repulsion would not be strongly sufficient to overcome the gravitational attraction in order to form the scalar hair. But, including the GB term with the negative GB coupling parameter $\alpha$ makes the spacetime curvature or the gravitational attraction weaker compared to Einstein gravity. This can be seen from the behavior of the effective asymptotic AdS radius $l_{\text{eff}}$ which is a decreasing function of $\alpha$. Hence, with $\alpha$ being negative and its magnitude being large enough, the electrostatic repulsion is easier to overcome the gravitational attraction, which results in the condensation of the scalar field around the event horizon of the planar black hole or the formation of Cooper pairs of quarks at the boundary field theory. As a result, the presence of the GB term with $\alpha<0$ can lead to the occurrence of the CSC phase for $N_c\geq2$. Second, is it possible to understand from the viewpoint of QCD? In the viewpoint of QCD, the CSC phase is realized as quarks condense into the Cooper pairs at sufficiently high chemical potential and low temperature. In this sense, one can understand that the presence of the GB term with $\alpha$ being negative and its magnitude being large enough would lead to the existence of a sufficiently high chemical potential (and low temperature) region, where the CSC phase (dual to the black hole with the scalar hair) lives, which is not found in Einstein gravity. Third, we observe that the occurrence of the CSC phase for $N_c\geq2$ requires the magnitude of the GB coupling parameter $\alpha$ to be quite large where the GB term would no longer be considered as the correction or in other words in this situation the GB term becomes important. Hence, the terms with the further powers of the curvature tensors such as $R^3$ or $R^4$ cannot be ignored but must be taken into account. In addition, the large magnitude of $\alpha$ violates the causality bound. These problems shall be resolved in the next section.
\subsection{Confinement phase}

\begin{figure}[t]
 \centering
\begin{tabular}{cc}
\includegraphics[width=0.6 \textwidth]{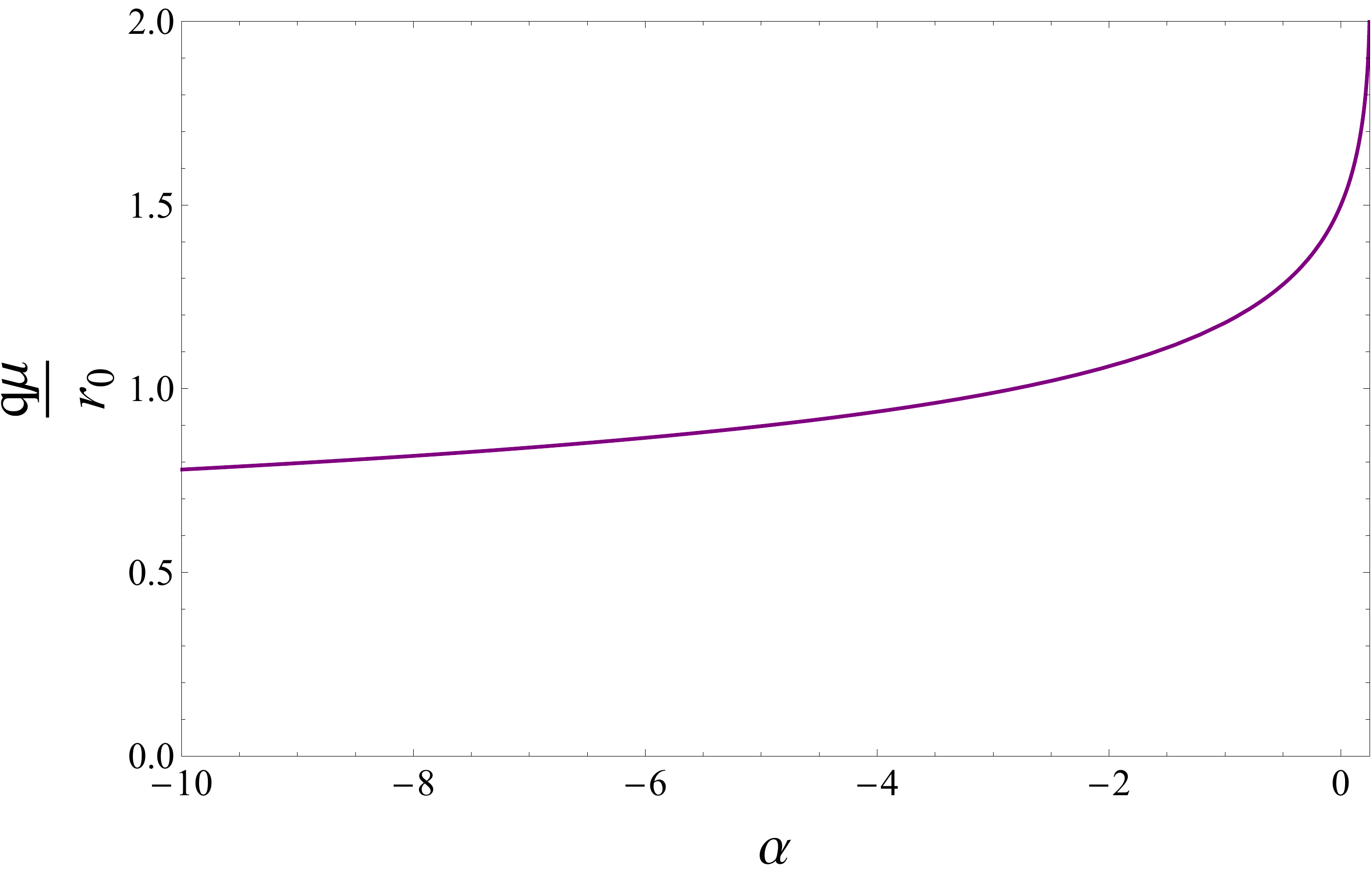}
\end{tabular}
 \caption{The depenedence of $q\mu/r_0$ in terms of the GB coupling parameter $\alpha$.}\label{CPBFb}
\end{figure}

In order to find the condensation of the scalar field in the confinement phase, we need to solve Eqs. (\ref{Sol-phi-Eq}) and (\ref{Sol-psi-Eq}) in the background of the GB-AdS soliton. First, let us determine the necessary condition which corresponds to the breakdown of the BF bound as
\begin{eqnarray}
\frac{q^2\phi^2(r)}{r}>\frac{9}{4l^2_{\text{eff}}},
\end{eqnarray}
with $\phi(r)$ given in Eq. (\ref{VPSo}), which leads to
\begin{eqnarray}
\frac{q\mu}{r_0}>\frac{3}{2}\sqrt{\frac{1-\sqrt{1-4\alpha}}{2\alpha}}.
\end{eqnarray}
In the case of Einstein gravity, we derive $q\mu>1.5$ for $r_0=1$. The behavior of $q\mu/r_0$ as a function of the GB coupling parameter $\alpha$ is shown in Fig. \ref{CPBFb}.

The sufficient condition for the condensation of the scalar field in the confinement phase can be obtained by solving numerically Eqs. (\ref{Sol-phi-Eq}) and (\ref{Sol-psi-Eq}) in the GB-AdS soliton background. The corresponding numerical values of the rescaled critical chemical potential $q\mu_c$ are given in Table \ref{CPCCP}.
\begin{table}[!htp]
\centering
\begin{tabular}{c|c||c|c}
  \hline
  \hline
  $\alpha$ & $q\mu_c$ & $\alpha$ & $q\mu_c$ \\
  \hline
  $\ \ \ \ \ \ 0\ \ \ \ \ \ $ & $\ \ \ \ \ \ 3.05195 \ \ \ \ \ \ $ & $\ \ \ \ \ \ -6.5\ \ \ \ \ \ $ & $\ \ \ \ \ \ 1.82547 \ \ \ \ \ \ $\\
  \hline
  $\ \ \ \ \ \ -2.0\ \ \ \ \ \ $ & $\ \ \ \ \ \ 2.24992 \ \ \ \ \ \ $ & $\ \ \ \ \ \ -7.0\ \ \ \ \ \ $ & $\ \ \ \ \ \ 1.79786 \ \ \ \ \ \ $\\
  \hline
  $\ \ \ \ \ \ -2.4\ \ \ \ \ \ $ & $\ \ \ \ \ \ 2.18620 \ \ \ \ \ \ $ & $\ \ \ \ \ \ -7.5\ \ \ \ \ \ $ & $\ \ \ \ \ \ 1.77611 \ \ \ \ \ \ $\\
  \hline
  $\ \ \ \ \ \ -2.8\ \ \ \ \ \ $ & $\ \ \ \ \ \ 2.13102 \ \ \ \ \ \ $ & $\ \ \ \ \ \ -8.0\ \ \ \ \ \ $ & $\ \ \ \ \ \ 1.75182 \ \ \ \ \ \ $\\
  \hline
  $\ \ \ \ \ \ -3.2\ \ \ \ \ \ $ & $\ \ \ \ \ \ 2.08212 \ \ \ \ \ \ $ & $\ \ \ \ \ \ -8.5\ \ \ \ \ \ $ & $\ \ \ \ \ \ 1.73558 \ \ \ \ \ \ $\\
  \hline
  $\ \ \ \ \ \ -3.8\ \ \ \ \ \ $ & $\ \ \ \ \ \ 2.01559 \ \ \ \ \ \ $ & $\ \ \ \ \ \ -9.0\ \ \ \ \ \ $ & $\ \ \ \ \ \ 1.71074 \ \ \ \ \ \ $\\
  \hline
  $\ \ \ \ \ \ -4.2\ \ \ \ \ \ $ & $\ \ \ \ \ \ 1.98261 \ \ \ \ \ \ $ & $\ \ \ \ \ \ -9.5\ \ \ \ \ \ $ & $\ \ \ \ \ \ 1.69185 \ \ \ \ \ \ $\\
  \hline
  \hline
\end{tabular}
\caption{The numerical values for $q\mu_c$ with various values of $\alpha$ in the confinement phase.} \label{CPCCP}
\end{table}
With $\alpha=-3.2$ and $\alpha=-2.4$, the values of the critical chemical potential $\mu_c$ are $2.08212$ and $2.18620$, respectively, for $N_c=2$. Because the confinement phase exists at the chemical potential which is below $1.73$, the CSC phase does not exist in the confinement phase with these values of $\alpha$, as seen in the top panels of Fig. \ref{PD-Nc23}. This happens similarly to the case of $N_c=3$ with $\alpha=-9.0$ and $\alpha=-8.0$. In addition, from Table \ref{CPCCP} we find that the scaled critical chemical potential $q\mu_c$ decreases with increasing the magnitude of the GB coupling parameter $\alpha$. This implies that for the sufficiently large magnitude of $\alpha$, the critical chemical potential $\mu_c$ would be lower than $1.73$ and since the CSC phase can appear even in the confinement phase. This may indicate the breakdown region of the GB term in investigating the CSC phase transition.

\section{\label{HCSC2} EGB holographic CSC with the additional corrections from matter}

In the previous section, we have indicated that the higher curvature corrections written as the GB term can lead to the CSC phase for $N_c\geq2$ with the appropriate value of the GB coupling parameter $\alpha$. However, in order to obtain the CSC phase for $N_c\geq2$, the GB coupling parameter $\alpha$ should be negative and its magnitude is rather large, which is beyond the region of the classical gravity and violates the causality bound. In this section, we will resolve this problem by considering additionally the higher derivative corrections from the matter and the non-minimal coupled Maxwell field.

The action of the system under the consideration is given by
\begin{equation}
S_{\text{bulk}}=\frac{1}{2\kappa^2_6}\int
d^6x\sqrt{-g}\left[R-2\Lambda+\widetilde{\alpha}\mathcal{L}_{GB}+\mathcal{L}_{\text{mat}}+\beta\mathcal{L}_{\text{non-min}}\right],\label{EGB-ED-adS-new}
\end{equation}
where the matter Lagrangian $\mathcal{L}_{\text{mat}}$ and the Lagrangian $\mathcal{L}_{\text{non-min}}$ describing the non-minimal coupled Maxwell field are
\begin{eqnarray}
\mathcal{L}_{\text{mat}}&=&-\frac{1}{4}F_{\mu\nu}F^{\mu\nu}+b\left(F_{\mu\nu}F^{\mu\nu}\right)^2+\mathcal{O}(b^2)-|(\nabla_\mu-iqA_\mu)\psi|^2-m^2|\psi|^2,\nonumber\\
\mathcal{L}_{\text{non-min}}&=&R\left(F_{\mu\nu}F^{\mu\nu}\right)^2-4R_{\mu\nu}F^{\mu\rho}{F^{\nu}}_\rho+R_{\mu\nu\rho\lambda}F^{\mu\nu}F^{\rho\lambda},
\end{eqnarray}                                                                                                                                                                                                                                                                                                                                                                                                                                                                                                                                                                                                                                                                                                                                                                                                                                                                                                                                                                                                                                                                                                                                                                                                                                                                                                                                                                                                                                                                                                                                                                                                                                                                                                                                                                                                                                                                                                                                                                                                                                                                                                                                                                                                                                                                                                                                                                                                                  
and $b$ and $\beta$ are the parameters characterizing the higher derivative correction for the Maxwell electrodynamics and the non-minimal coupled Maxwell field, respectively. Let us clarify the additionally higher derivative corrections in the action (\ref{EGB-ED-adS-new}). First, we have considered the higher derivative correction for the Maxwell electrodynamics at the first order described by the term $b\left(F_{\mu\nu}F^{\mu\nu}\right)^2$. Choosing this term is motivated by at least three reasons in order:
\begin{itemize}
\item[1.] Besides the GB term, the next-to-leading order corrections to the bosonic sector in the effective action of the heterotic string theory lead to the term in the form $\left(F_{\mu\nu}F^{\mu\nu}\right)^2$ \cite{Kats2007,Cai2008,JTLiu2009,Anninos2009}.

\item[2.] This term has been obtained as the one-loop correction of quantum electrodynamics (QED) \cite{Ritz1996}.

\item[3.] This term can be obtained as the next-to-leading order correction in the expansion of the Born-Infeld-type electrodynamics which are well-known as Born-Infeld, logarithmic and exponential electrodynamics \cite{Hendi2012}.
\end{itemize}
Second, $\mathcal{L}_{\text{non-min}}$ which describes the non-minimal coupled Maxwell field is constructed from the Riemann tensor and the Maxwell field strength tensor \cite{XHFeng2016}. The motivation for considering the non-minimal coupling in this form is the equations of motion for the metric field and the Maxwell potential remain the second order in the derivatives. Furthermore, we have not considered the higher derivative corrections corresponding to the scalar field. This is because we investigate the system near the critical chemical potential where the value of the scalar field is near zero and hence including the corresponding corrections does not lead to the significant effects.

From the viewpoint of string theory, we can expect that the stringy corrections to leading order to Einstein gravity coupled to the $U(1)$ gauge field would include all possible four-derivative terms \cite{Natsuume1994,Sachdev2011}. The four-derivative terms are constructed from the metric curvatures, the field strength tensor of the $U(1)$ gauge field and its derivatives. Interestingly, by making a field redefinition of the following general form
\begin{eqnarray}
g_{\mu\nu}&\longrightarrow& g_{\mu\nu}+a_1R_{\mu\nu}+a_2F_{\mu\rho}{F_\nu}^\rho+\left(a_3R+a_4F_{\rho\sigma}F^{\rho\sigma}\right)g_{\mu\nu},\nonumber\\
A_\mu&\longrightarrow&A_\mu+\lambda\nabla^\nu F_{\mu\nu},\label{field-ref}
\end{eqnarray}
where $a_i$ and $\lambda$ are constants, one can reorganize all four-derivative terms to make the calculations more simple. Under the field redefinition (\ref{field-ref}), the coupling parameters associated with the four-derivative terms are changed, where their change is to depend on $a_i$ and $\lambda$ and is explicitly given in Ref. \cite{Natsuume1994}. As a result, with appropriate constants $a_i$ and $\lambda$, we can assemble the curvature-squared terms into the GB term and the terms of coupling between the metric curvatures and the field strength tensor of the $U(1)$ gauge field into $\mathcal{L}_{\text{non-min}}$. In addition, for the radial electric field configuration of the $U(1)$ gauge field where only the components of $F^{01}$ and $F^{10}$ are non-zero, the term $F^{\mu\nu}F_{\nu\rho}F^{\rho\sigma}F_{\sigma\mu}$ contributes to the equation of motion to be equal to half the contribution of the term $\left(F_{\mu\nu}F^{\mu\nu}\right)^2$. This means that the role of $F^{\mu\nu}F_{\nu\rho}F^{\rho\sigma}F_{\sigma\mu}$ and $\left(F_{\mu\nu}F^{\mu\nu}\right)^2$ in this situation is the same and since their effects on the CSC phase transition can be characterized in terms of a unique parameter $b$.
 
The equations of motion corresponding to the action (\ref{EGB-ED-adS-new}) are found as
\begin{eqnarray}
G_{\mu\nu}+\widetilde{\alpha}H_{\mu\nu}-\frac{10}{l^2}g_{\mu\nu}&=&T_{\mu\nu},\nonumber\\
\nabla_\mu\left[\left(1-8b F_{\rho\sigma}F^{\rho\sigma}\right)F^{\mu\nu}-\beta\delta^{\rho\lambda\mu\nu}_{\rho'\lambda'\sigma\gamma}{R^{\rho'\lambda'}}_{\rho\lambda}F^{\sigma\gamma}\right]&=&iq\left[\psi^*(\nabla^\nu-iqA^\nu)\psi-\psi(\nabla^\nu+iqA^\nu)\psi^*\right],\nonumber\\
(\nabla_\mu-iqA_\mu)(\nabla^\mu-iqA^\mu)\psi-m^2\psi&=&0,\label{EOM2}
\end{eqnarray}
where
\begin{eqnarray}
T_{\mu\nu}&=&\frac{1}{2}\left(1-8b F_{\rho\sigma}F^{\rho\sigma}\right)F_{\mu\lambda}{F_\nu}^\lambda-\frac{\beta}{2}\left[\delta^{\rho\lambda\sigma\gamma}_{\rho'\lambda'\sigma'\mu}{R^{\rho'\lambda'}}_{\rho\lambda}F_{\sigma\gamma}{F^{\sigma'}}_\nu+\frac{1}{2}\delta^{\rho\lambda\sigma\gamma}_{\rho'\mu\lambda'\sigma'}{R^{\rho'}}_{\nu\rho\lambda}F^{\lambda'\sigma'}F_{\sigma\gamma}\right.\nonumber\\
&&\left.+g_{\rho\mu}\delta^{\rho\lambda\sigma\gamma}_{\nu\rho'\lambda'\sigma'}\nabla^{\rho'}\nabla_\lambda\left(F^{\lambda'\sigma'}F_{\sigma\gamma}\right)\right]+\frac{1}{2}\left[(\nabla_\nu-iqA_\nu)\psi(\nabla_\mu+iqA_\mu)\psi^*+\mu\leftrightarrow\nu\right]\nonumber\\
&&+\frac{1}{2}g_{\mu\nu}\left(\mathcal{L}_{\text{mat}}+\beta\mathcal{L}_{\text{non-min}}\right),
\end{eqnarray}
and $\delta^{\mu\nu\rho\lambda}_{\mu'\nu'\rho'\lambda'}$ is the generalized Kronecker delta function which is totally antisymmetric in the upper indices as well as the lower indices.

First, we shall show that only considering additionally the higher derivative correction for the Maxwell electrodynamics can provide a suitable holographic model of the CSC phase transition where the GB and $b\left(F_{\mu\nu}F^{\mu\nu}\right)^2$ terms actually play the role of the corrections which means the parameters $\alpha$ and $b$ to be small. Thus, let us turn off the non-minimal coupled term, i.e. $\beta=0$, in the following analyses. 

In the deconfinement phase, the equations for $f(r)$ and $\phi(r)$ are found as
\begin{eqnarray}
\alpha\left[2f'(r)r+5f(r)\right]f(r)-rf'(r)-5f(r)+5-\frac{1}{8}\phi'(r)^2\left[1+24b\phi'(r)^2\right]&=&0,\label{r-f-Eq-mod}\\
\phi''(r)+\frac{4}{r}\left[1-32b\phi'(r)^2\right]\phi'(r)-\frac{2q^2\psi^2(r)}{r^2f(r)}\phi(r)\left[1-48b\phi'(r)^2\right]&=&0.\label{r-phi-Eq-mod}
\end{eqnarray}
Note that, the equation for $\psi(r)$ does not get modified when including the higher derivative correction for the Maxwell electrodynamics and it is given by Eq. (\ref{r-psi-Eq}). Near the critical chemical potential $\mu_c$ where the backreaction of the scalar field is negligible and at the first order coming from the higher derivative correction for the Maxwell electrodynamics, we find 
\begin{eqnarray}
f(r)&=&\frac{1}{2\alpha}\left\{1-\left[1-4\alpha\left(1-\frac{r^5_+}{r^5}\right)+\frac{3\alpha\mu^2}{2r^2_+}\left(\frac{r_+}{r}\right)^5\left(1-\frac{r^3_+}{r^3}\right)\right.\right.\nonumber\\
&&\left.\left.+\frac{324b}{11}\frac{\alpha\mu^4}{r^4_+}\left(\frac{r_+}{r}\right)^5\left(3-4\frac{r^3_+}{r^3}+\frac{r^{11}_+}{r^{11}}\right)\right]^{1/2}\right\},\label{GBRNAdSfr-mod}\\
\phi(r)&=&\mu\left(1-\frac{r^3_+}{r^3}\right)-\frac{432b}{11}\frac{\mu^3r_+}{r^3}\left(1-\frac{r^8_+}{r^8}\right).\label{phi-r-mod}
\end{eqnarray}
In addition, we find the Hawking temperature and the free energy of the planar RN-AdS black hole of the higher derivative corrections as
\begin{eqnarray}
T&=&\frac{1}{4\pi}\left(5r_+-\frac{9\mu^2}{8r_+}-\frac{81b\mu^4}{11r^3_+}\right),\nonumber\\
\Omega_{\text{BH}}&=&-r^5_+\left(1+\frac{3\mu^2}{8r^2_+}+\frac{1215b\mu^4}{11r^4_+}\right)\frac{4\pi}{5r_0}V_3.
\end{eqnarray}

\begin{figure}[t]
 \centering
\begin{tabular}{cc}
\includegraphics[width=0.5 \textwidth]{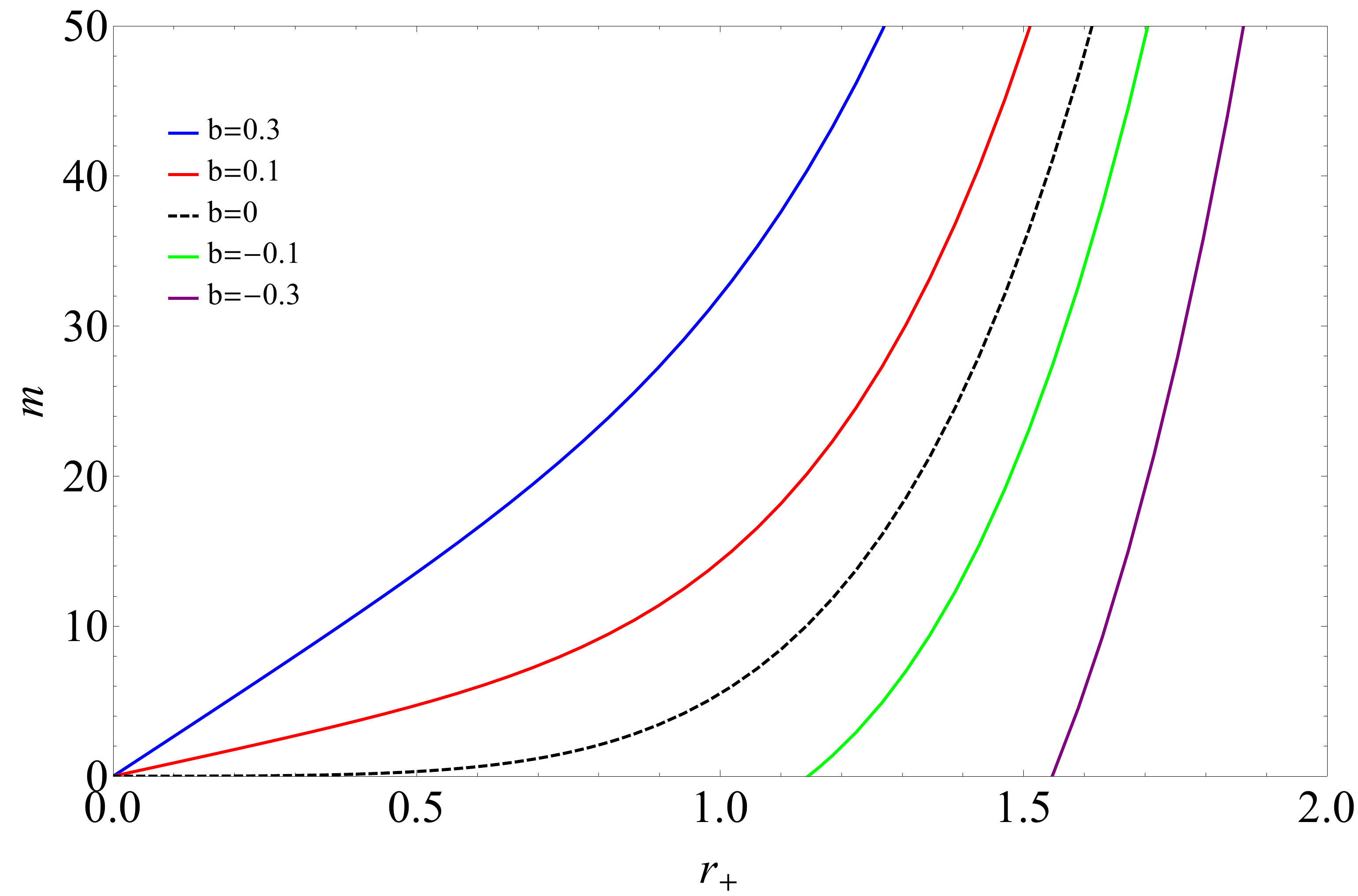}
\end{tabular}
 \caption{The scaled ADM mass $m$ of the black hole as a function in terms of the event horizon radius $r_+$ for various values of the parameter $b$. For $m$ kept fixed, the event horizon radius increases with decreasing the parameter $b$, which suggests the weaker gravitational attraction (around the event horizon) for the lower $b$.}\label{dep-m-r+}
\end{figure}

By solving numerically Eqs. (\ref{r-f-Eq-mod}), (\ref{r-phi-Eq-mod}), and (\ref{r-psi-Eq}), we find the scaled critical chemical potential $\mu_c/r_+$ and the slope $T_c/\mu_c$ of the critical line $T_c=T_c(\mu_c)$. Their numerical results are shown in Tables \ref{Nc2-mod} ($N_c=2)$ and \ref{Nc3-mod} ($N_c=3$) for various values of the parameters $b$ and $\alpha$.
\begin{table}[!htp]
\centering
\begin{tabular}{c|c|c||c|c|c}
  \hline
  \hline
  \multicolumn{3}{c||}{$\alpha=-0.05$} & \multicolumn{3}{c}{$\alpha=-0.005$}\\
  \hline
  $b\times10^4$ & $\mu_c/r_+$ & $T_c/\mu_c$ & $b\times10^4$ & $\mu_c/r_+$ & $T_c/\mu_c$ \\
  \hline
  $\ \ \ \  -2.6 \ \ \ \ $ & $ \ \ \ \ 2.29828 \ \ \ \ $ & $\ \ \ \ 0.00531 \ \ \ \ $ &
  $ \ \ \ \ -2.6 \ \ \ \ $ & $ \ \ \ \ 2.29633 \ \ \ \ $ & $\ \ \ \ 0.00544 \ \ \ \ $\\ 
  \hline
  $\ \ \ \  -2.7 \ \ \ \ $ & $ \ \ \ \ 2.32662 \ \ \ \ $ & $\ \ \ \ 0.00792 \ \ \ \ $ &
  $ \ \ \ \ -2.7 \ \ \ \ $ & $ \ \ \ \ 2.32456 \ \ \ \ $ & $\ \ \ \ 0.00800 \ \ \ \ $\\
  \hline
  $\ \ \ \  -2.8 \ \ \ \ $ & $ \ \ \ \ 2.36011 \ \ \ \ $ & $\ \ \ \ 0.01185 \ \ \ \ $ &
  $ \ \ \ \ -2.8 \ \ \ \ $ & $ \ \ \ \ 2.35814 \ \ \ \ $ & $\ \ \ \ 0.01188 \ \ \ \ $\\
  \hline
  $\ \ \ \  -2.9 \ \ \ \ $ & $ \ \ \ \ 2.40059 \ \ \ \ $ & $\ \ \ \ 0.01795 \ \ \ \ $ &
  $ \ \ \ \ -2.9 \ \ \ \ $ & $ \ \ \ \ 2.39912 \ \ \ \ $ & $\ \ \ \ 0.01791 \ \ \ \ $\\
  \hline
  $\ \ \ \  -3.0 \ \ \ \ $ & $ \ \ \ \ 2.45149 \ \ \ \ $ & $\ \ \ \ 0.02788 \ \ \ \ $ &
  $ \ \ \ \ -3.0 \ \ \ \ $ & $ \ \ \ \ 2.45145 \ \ \ \ $ & $\ \ \ \ 0.02787 \ \ \ \ $\\
  \hline
  $\ \ \ \  -3.1 \ \ \ \ $ & $ \ \ \ \ 2.52066 \ \ \ \ $ & $\ \ \ \ 0.04580 \ \ \ \ $ &
  $ \ \ \ \ -3.1 \ \ \ \ $ & $ \ \ \ \ 2.52491 \ \ \ \ $ & $\ \ \ \ 0.04651 \ \ \ \ $\\
  \hline
  \hline
  \multicolumn{3}{c||}{$\alpha=0.001$} & \multicolumn{3}{c}{$\alpha=0.01$}\\
  \hline
  $b\times10^4$ & $\mu_c/r_+$ & $T_c/\mu_c$ & $b\times10^4$ & $\mu_c/r_+$ & $T_c/\mu_c$ \\
  \hline
  $\ \ \ \  -2.6 \ \ \ \ $ & $ \ \ \ \ 2.29600 \ \ \ \ $ & $\ \ \ \ 0.00546 \ \ \ \ $ &
  $ \ \ \ \ -2.6 \ \ \ \ $ & $ \ \ \ \ 2.29548 \ \ \ \ $ & $\ \ \ \ 0.00549 \ \ \ \ $\\
  \hline
  $\ \ \ \  -2.7 \ \ \ \ $ & $ \ \ \ \ 2.32420 \ \ \ \ $ & $\ \ \ \ 0.00802 \ \ \ \ $ &
  $ \ \ \ \ -2.7 \ \ \ \ $ & $ \ \ \ \ 2.32362 \ \ \ \ $ & $\ \ \ \ 0.00804 \ \ \ \ $\\
  \hline
  $\ \ \ \  -2.8 \ \ \ \ $ & $ \ \ \ \ 2.35777 \ \ \ \ $ & $\ \ \ \ 0.01188 \ \ \ \ $ &
  $ \ \ \ \ -2.8 \ \ \ \ $ & $ \ \ \ \ 2.35715 \ \ \ \ $ & $\ \ \ \ 0.01189 \ \ \ \ $\\
  \hline
  $\ \ \ \  -2.9 \ \ \ \ $ & $ \ \ \ \ 2.39878 \ \ \ \ $ & $\ \ \ \ 0.01790 \ \ \ \ $ &
  $ \ \ \ \ -2.9 \ \ \ \ $ & $ \ \ \ \ 2.39818 \ \ \ \ $ & $\ \ \ \ 0.01789 \ \ \ \ $\\
  \hline
  $\ \ \ \  -3.0 \ \ \ \ $ & $ \ \ \ \ 2.45124 \ \ \ \ $ & $\ \ \ \ 0.02786 \ \ \ \ $ &
  $ \ \ \ \ -3.0 \ \ \ \ $ & $ \ \ \ \ 2.45082 \ \ \ \ $ & $\ \ \ \ 0.02782 \ \ \ \ $\\
  \hline
  $\ \ \ \  -3.1 \ \ \ \ $ & $ \ \ \ \ 2.52519 \ \ \ \ $ & $\ \ \ \ 0.04656 \ \ \ \ $ &
  $ \ \ \ \ -3.1 \ \ \ \ $ & $ \ \ \ \ 2.52545 \ \ \ \ $ & $\ \ \ \ 0.04660 \ \ \ \ $\\ 
  \hline
  \hline
\end{tabular}
\caption{The numerical values for $\mu_c/r_+$ and $T_c/\mu_c$ with various values of $b$ and $\alpha$ at $N_c=2$.} \label{Nc2-mod}
\end{table}
\begin{table}[!htp]
\centering
\begin{tabular}{c|c|c||c|c|c}
  \hline
  \hline
  \multicolumn{3}{c||}{$\alpha=-0.01$} & \multicolumn{3}{c}{$\alpha=-0.001$}\\
  \hline
  $b\times10^4$ & $\mu_c/r_+$ & $T_c/\mu_c$ & $b\times10^4$ & $\mu_c/r_+$ & $T_c/\mu_c$ \\
  \hline
  $\ \ \ \  -2.70 \ \ \ \ $ & $ \ \ \ \ 2.41668 \ \ \ \ $ & $\ \ \ \ 0.00585 \ \ \ \ $ &
  $ \ \ \ \ -2.70 \ \ \ \ $ & $ \ \ \ \ 2.41536 \ \ \ \ $ & $\ \ \ \ 0.00585 \ \ \ \ $\\
  \hline
  $\ \ \ \  -2.73 \ \ \ \ $ & $ \ \ \ \ 2.44244 \ \ \ \ $ & $\ \ \ \ 0.00780 \ \ \ \ $ &
  $ \ \ \ \ -2.73 \ \ \ \ $ & $ \ \ \ \ 2.44081 \ \ \ \ $ & $\ \ \ \ 0.00778 \ \ \ \ $\\
  \hline
  $\ \ \ \  -2.76 \ \ \ \ $ & $ \ \ \ \ 2.47361 \ \ \ \ $ & $\ \ \ \ 0.01062 \ \ \ \ $ &
  $ \ \ \ \ -2.76 \ \ \ \ $ & $ \ \ \ \ 2.47152 \ \ \ \ $ & $\ \ \ \ 0.01054 \ \ \ \ $\\
  \hline
  $\ \ \ \  -2.79 \ \ \ \ $ & $ \ \ \ \ 2.51377 \ \ \ \ $ & $\ \ \ \ 0.01500 \ \ \ \ $ &
  $ \ \ \ \ -2.79 \ \ \ \ $ & $ \ \ \ \ 2.51088 \ \ \ \ $ & $\ \ \ \ 0.01481 \ \ \ \ $\\
  \hline
  $\ \ \ \  -2.82 \ \ \ \ $ & $ \ \ \ \ 2.57424 \ \ \ \ $ & $\ \ \ \ 0.02326 \ \ \ \ $ &
  $ \ \ \ \ -2.82 \ \ \ \ $ & $ \ \ \ \ 2.56926 \ \ \ \ $ & $\ \ \ \ 0.02269 \ \ \ \ $\\
  \hline
  $\ \ \ \  -2.85 \ \ \ \ $ & $ \ \ \ \ 2.68366 \ \ \ \ $ & $\ \ \ \ 0.04525 \ \ \ \ $ &
  $ \ \ \ \ -2.85 \ \ \ \ $ & $ \ \ \ \ 2.68291 \ \ \ \ $ & $\ \ \ \ 0.04508 \ \ \ \ $\\
  \hline
  \hline
  \multicolumn{3}{c||}{$\alpha=0.0001$} & \multicolumn{3}{c}{$\alpha=0.001$}\\
  \hline
  $b\times10^4$ & $\mu_c/r_+$ & $T_c/\mu_c$ & $b\times10^4$ & $\mu_c/r_+$ & $T_c/\mu_c$ \\
  \hline
  $\ \ \ \  -2.70 \ \ \ \ $ & $ \ \ \ \ 2.41520 \ \ \ \ $ & $\ \ \ \ 0.00585 \ \ \ \ $ &
  $ \ \ \ \ -2.70 \ \ \ \ $ & $ \ \ \ \ 2.41506 \ \ \ \ $ & $\ \ \ \ 0.00585 \ \ \ \ $\\
  \hline
  $\ \ \ \  -2.73 \ \ \ \ $ & $ \ \ \ \ 2.44061 \ \ \ \ $ & $\ \ \ \ 0.00778 \ \ \ \ $ &
  $ \ \ \ \ -2.73 \ \ \ \ $ & $ \ \ \ \ 2.44044 \ \ \ \ $ & $\ \ \ \ 0.00778 \ \ \ \ $\\
  \hline
  $\ \ \ \  -2.76 \ \ \ \ $ & $ \ \ \ \ 2.47126 \ \ \ \ $ & $\ \ \ \ 0.01053 \ \ \ \ $ &
  $ \ \ \ \ -2.76 \ \ \ \ $ & $ \ \ \ \ 2.47104 \ \ \ \ $ & $\ \ \ \ 0.01052 \ \ \ \ $\\
  \hline
  $\ \ \ \  -2.79 \ \ \ \ $ & $ \ \ \ \ 2.51052 \ \ \ \ $ & $\ \ \ \ 0.01478 \ \ \ \ $ &
  $ \ \ \ \ -2.79 \ \ \ \ $ & $ \ \ \ \ 2.51023 \ \ \ \ $ & $\ \ \ \ 0.01476 \ \ \ \ $\\
  \hline
  $\ \ \ \  -2.82 \ \ \ \ $ & $ \ \ \ \ 2.56865 \ \ \ \ $ & $\ \ \ \ 0.02262 \ \ \ \ $ &
  $ \ \ \ \ -2.82 \ \ \ \ $ & $ \ \ \ \ 2.56815 \ \ \ \ $ & $\ \ \ \ 0.02257 \ \ \ \ $\\
  \hline
  $\ \ \ \  -2.85 \ \ \ \ $ & $ \ \ \ \ 2.68283 \ \ \ \ $ & $\ \ \ \ 0.04506 \ \ \ \ $ &
  $ \ \ \ \ -2.85 \ \ \ \ $ & $ \ \ \ \ 2.68275 \ \ \ \ $ & $\ \ \ \ 0.04504 \ \ \ \ $\\
  \hline
  \hline
\end{tabular}
\caption{The numerical values for $\mu_c/r_+$ and $T_c/\mu_c$ with various values of $b$ and $\alpha$ at $N_c=3$.} \label{Nc3-mod}
\end{table}
From these tables, we observe that the CSC phase transition with $N_c\geq2$ can be achieved with including the higher derivative corrections for both Einstein gravity and Maxwell electrodynamics where the GB coupling parameter $\alpha$ is either positive or negative and satisfies the causality bound. Furthermore, we find that the GB coupling parameter $\alpha$ is small or in other words $\alpha$ belongs the perturbative region. This is clearly consistent to the fact that the GB term is considered as the correction and hence the further powers of the curvature tensors such as $R^3$ or $R^4$ are small compared to the GB term and hence they can be ignored. In particular, the sign of the parameter $b$ characterizing the higher derivative correction at the first order for the Maxwell electrodynamics should be negative in order to make the condensation of the scalar field appearing around the event horizon $r_+$. This can be explained as follows: as we discussed above, the charge $q=2/N_c$ of the dual scalar field decreases as increasing the color number and thus the electrostatic repulsion would not be strongly sufficient to overcome the gravitational attraction for the scalar field to be condensed; the presence of the GB term can make the gravitational attraction weaker if $\alpha$ is negative and its magnitude is large enough; however, by including the higher derivative correction at the first order for the Maxwell electrodynamics with the negative value of $b$, the event horizon of the black hole becomes larger as seen in Fig. \ref{dep-m-r+} and consequently it leads the sufficiently weak gravitational attraction around the event horizon such that the scalar hair can be formed even the small magnitude of the GB coupling parameter $\alpha$.

In particular, we see from Tables \ref{Nc2-mod} and \ref{Nc3-mod} that the CSC phase transition occurs for the case of the positive GB coupling parameter $\alpha>0$ (which is still true when the non-minimal coupled Maxwell field is turned on as seen later), which is not realized in the framework of the pure EGB gravity. This case is suitable to string theory, because the GB term arises naturally from the low-energy effective action of heterotic string theory at the order $\alpha'$ \cite{Zwiebach1985,Witten1986,Gross1987,Tseytlin1987,Bento1996} where the GB coupling parameter is regarded as the inverse string tension and thus is larger than zero. Furthermore, the CSC phase transition with $\alpha>0$ satisfies the constraint imposed by
the weak gravity conjecture on the GB coupling parameter. It was indicated in Ref. \cite{Kats2007} that the weak gravity conjecture excludes the entire region $\alpha<0$ for the pure EGB gravity.

\begin{figure}[t]
 \centering
\begin{tabular}{cc}
\includegraphics[width=0.45 \textwidth]{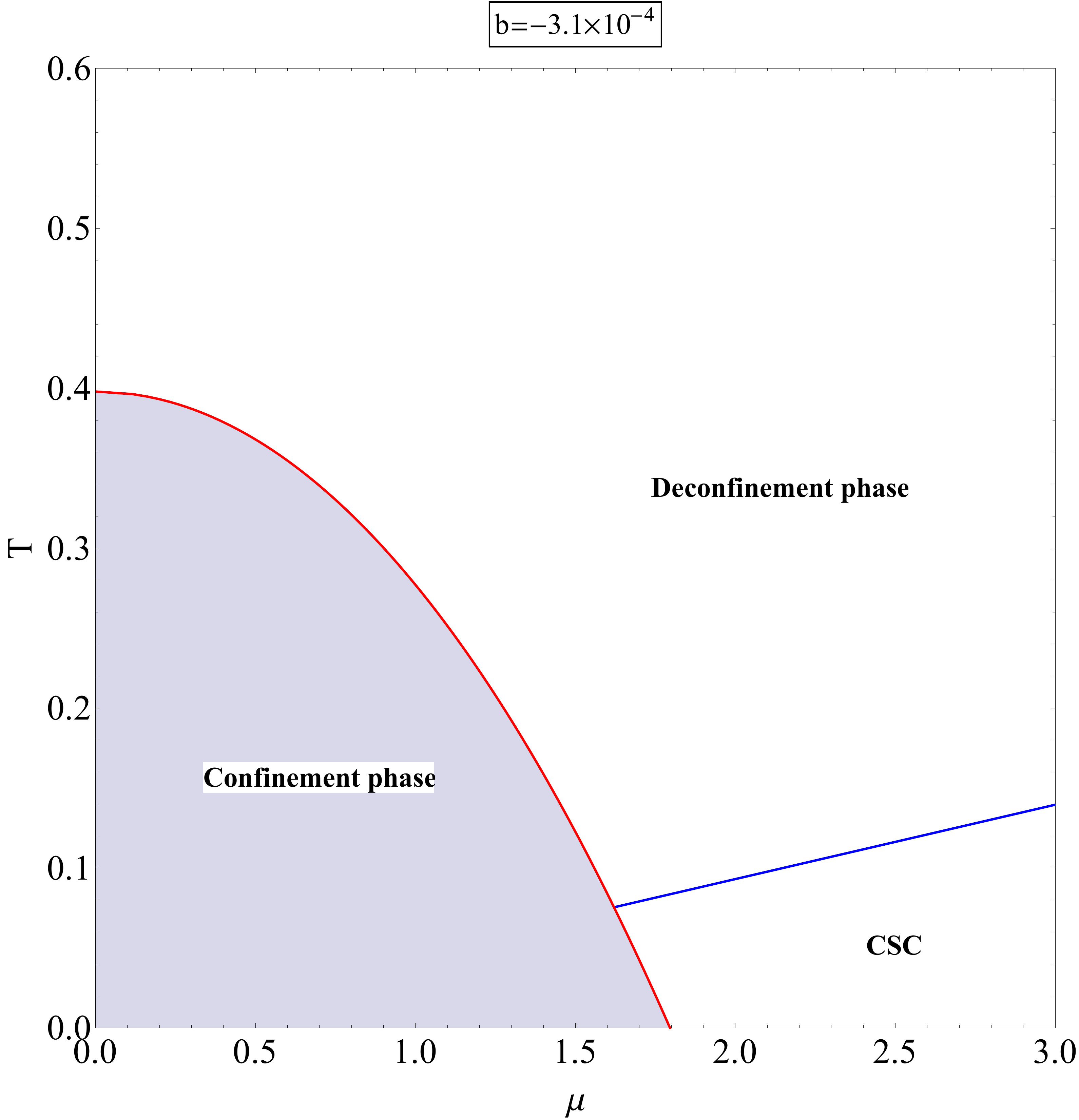}
\hspace*{0.05\textwidth}
\includegraphics[width=0.45 \textwidth]{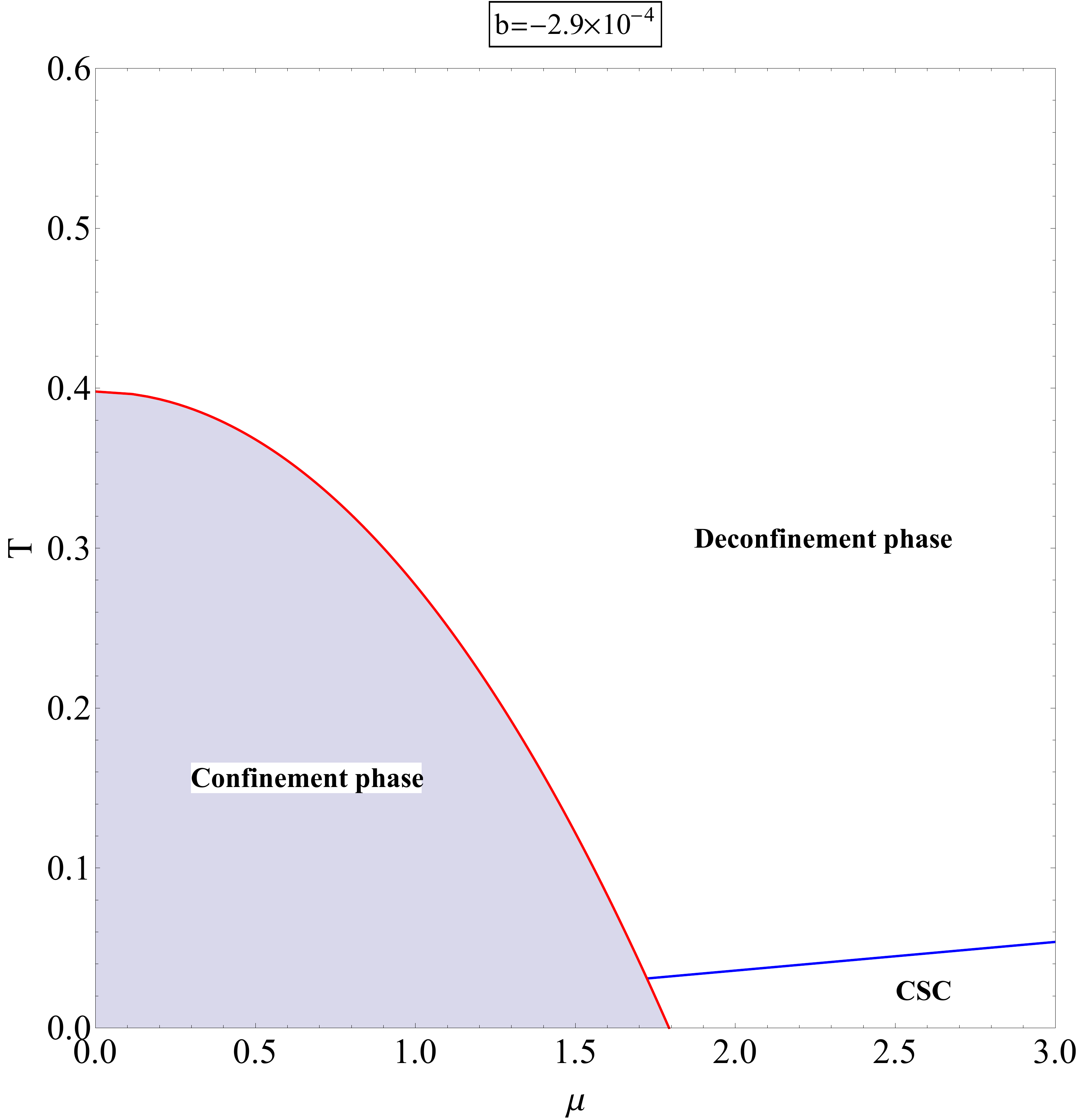}\\
\includegraphics[width=0.45 \textwidth]{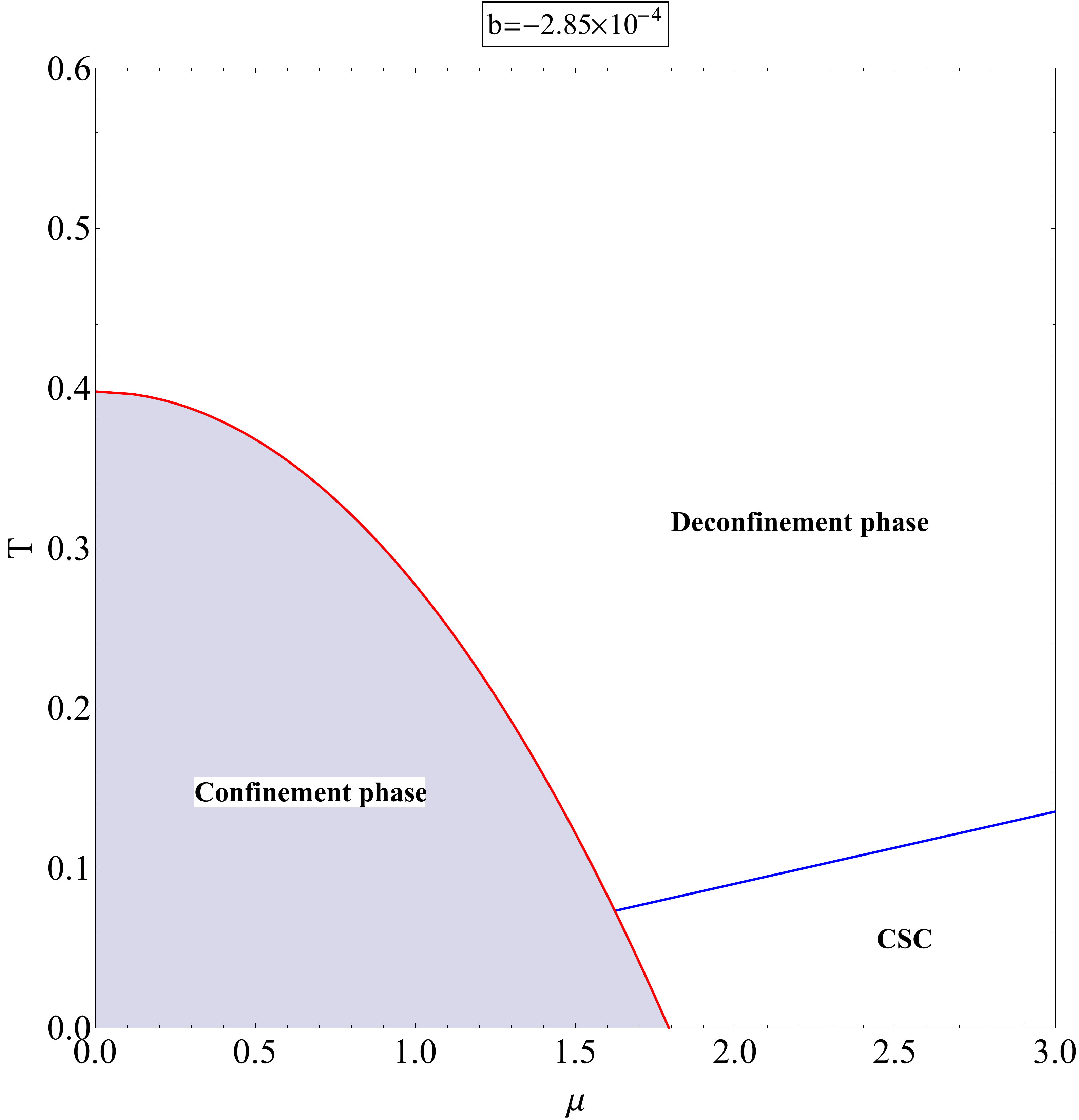}
\hspace*{0.05\textwidth}
\includegraphics[width=0.45 \textwidth]{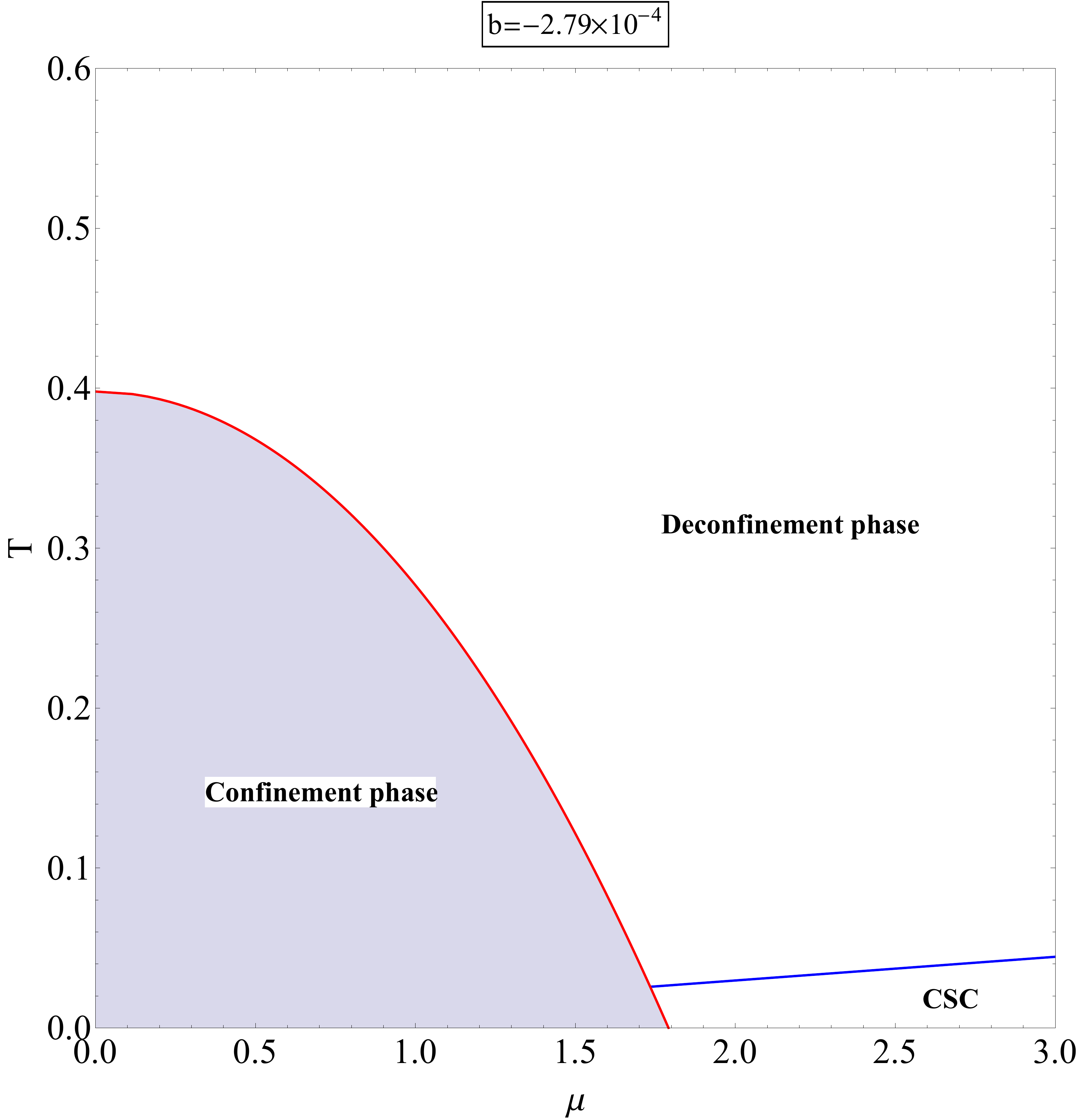}
\end{tabular}
  \caption{The phase diagram for various values of $b$. The values $N_c=2$ and $\alpha=-0.005$ are used for the top panels, whereas $N_c=3$ and $\alpha=-0.001$ are used for the bottom panels.}\label{PDmod-Nc23}
\end{figure}

In Fig. \ref{PDmod-Nc23}, we show the phase diagram of the YM theory with the presence of the CSC phase whose holographic model is EGB gravity coupled minimally to the vector and scalar fields with including the higher derivative correction for the Maxwell electrodynamics.
We find that the region corresponding to the CSC phase is larger and thus the CSC phase is more stable as increasing the magnitude of the parameter $b$. This can be seen from Tables \ref{Nc2-mod} and \ref{Nc3-mod} where the slope $T_c/\mu_c$ of the critical line $T_c=T_c(\mu_c)$ increases with the growth of the magnitude of the parameter $b$. Also, there is no the CSC state in the confinement phase for the parameters considered in Fig. \ref{PDmod-Nc23} because by solving numerically the equations for $\phi(r)$ and $\psi(r)$ in the confinement phase, we obtain the values of the critical chemical potential as $3.06014$ (with $\alpha=-0.005$ and $b\approx-3.0\times10^{-4}$) and $4.57641$ (with $\alpha=-0.001$ and $b\approx-2.8\times10^{-4}$) for $N_c=2$ and $N_c=3$, respectively, which are out of the confinement phase. Here, the equations for $\phi(r)$ in the confinement background is given by
\begin{eqnarray}
\phi''(r)+\left\{\frac{4}{r}\left[1-32bf(r)^2\phi'(r)^2\right]+\frac{f'(r)}{f(r)}\right\}\phi'(r)-\frac{2q^2\psi^2(r)}{r^2f(r)}\phi(r)\left[1-48bf(r)^2\phi'(r)^2\right]&=&0,\label{Sol-phi-Eq-mod}
\end{eqnarray}
where the expression for $f(r)$ is given in Eq. (\ref{fr-soliton}), whereas, the equation for $\psi(r)$ in the same background is given in Eq. (\ref{Sol-psi-Eq}).

Now we study the effect of the non-minimal coupled Maxwell field on the CSC phase transition by turning on the Lagrangian $\mathcal{L}_{\text{non-min}}$, i.e. $\beta\neq0$. In this case, the planar black hole dual to the deconfinement phase is described by the following ansatz
\begin{eqnarray}
ds^2=r^2\left(-f(r)e^{-\chi(r)}dt^2+h_{ij}dx^idx^j+dy^2\right)+\frac{dr^2}{r^2f(r)}.\label{BHa-new}
\end{eqnarray} 
The corresponding Hawking temperature is given by
\begin{eqnarray}
T=\frac{r^2_+f'(r_+)}{4\pi}e^{-\chi(r_+)/2}.
\end{eqnarray}

By substituting the ansatz (\ref{BHa-new}) for the metric, and (\ref{vec-scal-ans}) for the vector and scalar fields into Eq. (\ref{EOM2}), we obtain the equations for $f(r)$, $\chi(r)$, $\phi(r)$, and $\psi(r)$ as
\begin{eqnarray}
\alpha\left[2f'(r)r+5f(r)\right]f(r)-rf'(r)-5f(r)+5-\frac{e^{\chi(r)}}{8}\phi'(r)^2&&\nonumber\\
\times\left[1+24be^{\chi(r)}\phi'(r)^2+48\beta f(r)\right]&=&0,\label{r-f-Eq-mod2}\\
\left[1-2\alpha f(r)\right]\chi'(r)-\frac{12\beta}{r}e^{\chi(r)}\phi'(r)^2&=&0,\label{r-chi-Eq}\\
\left[1-48\beta f(r)\right]\phi''(r)+\left[\frac{4-192\beta f(r)}{r}+\frac{1-48\beta f(r)}{2}\chi'(r)-48\beta f'(r)\right]\phi'(r)&&\nonumber\\
-\frac{128}{r}be^{\chi(r)}\phi'(r)^3-\frac{2q^2\psi^2(r)}{r^2f(r)}\phi(r)\left[1-48be^{\chi(r)}\phi'(r)^2\right]&=&0,\label{r-phi-Eq-mod2}\\
\psi''(r)+\left[\frac{f'(r)}{f(r)}+\frac{6}{r}-\frac{\chi'(r)}{2}\right]\psi'(r)+\frac{1}{r^2f(r)}\left[\frac{q^2e^{\chi(r)}\phi^2(r)}{r^2f(r)}-m^2\right]\psi(r)&=&0.\label{r-psi-Eq-mod2}
\end{eqnarray}
The function $\chi(r)$ is finite at the event horizon $r_+$ and it approaches zero near the AdS boundary ($r\rightarrow\infty$). Note that, for the parameter $\beta$ associated with the non-minimal coupled Maxwell field to be small, we solve analytically Eq. (\ref{r-chi-Eq}) up to the first order in $\beta$ as
\begin{eqnarray}
e^{-\chi(r)}=-12\beta\int\frac{\phi'(r)^2dr}{r\left[1-2\alpha f(r)\right]}+C,
\end{eqnarray}
where the constant $C$ is fixed such that $e^{-\chi(r)}$ approaches one in the limit $r\rightarrow\infty$, the functions $f(r)$ and $\phi(r)$ are given in Eqs. (\ref{GBRNAdSfr-mod}) and (\ref{phi-r-mod}), respectively.

In order to find the critical temperature for the CSC phase transition when the non-minimal coupled Maxwell field is included, we solve numerically Eqs. (\ref{r-f-Eq-mod2}) $-$ (\ref{r-psi-Eq-mod2}). The corresponding numerical results are given in Table \ref{Nc23-mod} for various values of the parameter $\beta$.
\begin{table}[!htp]
\centering
\begin{tabular}{c|c|c||c|c|c}
  \hline
  \hline
  \multicolumn{3}{c||}{$N_c=2$, $\alpha=-0.005$, $b=-3.0\times10^{-4}$} & \multicolumn{3}{c}{$N_c=2$, $\alpha=0.001$, $b=-3.1\times10^{-4}$}\\
  \hline
  $\beta\times10^3$ & $\mu_c/r_+$ & $T_c/\mu_c$ & $\beta\times10^3$ & $\mu_c/r_+$ & $T_c/\mu_c$ \\
  \hline
  $\ \ \ \  -1.0 \ \ \ \ $ & $ \ \ \ \ 2.41960 \ \ \ \ $ & $\ \ \ \ 0.03104 \ \ \ \ $ &
  $ \ \ \ \ -1.0 \ \ \ \ $ & $ \ \ \ \ 2.53725 \ \ \ \ $ & $\ \ \ \ 0.06416 \ \ \ \ $\\
  \hline
  $\ \ \ \  -0.5 \ \ \ \ $ & $ \ \ \ \ 2.43447 \ \ \ \ $ & $\ \ \ \ 0.02925 \ \ \ \ $ &
  $ \ \ \ \ -0.5 \ \ \ \ $ & $ \ \ \ \ 2.52409 \ \ \ \ $ & $\ \ \ \ 0.05291 \ \ \ \ $\\
  \hline
  $\ \ \ \  -0.1 \ \ \ \ $ & $ \ \ \ \ 2.44791 \ \ \ \ $ & $\ \ \ \ 0.02812 \ \ \ \ $ &
  $ \ \ \ \ -0.1 \ \ \ \ $ & $ \ \ \ \ 2.52429 \ \ \ \ $ & $\ \ \ \ 0.04761 \ \ \ \ $\\
  \hline
  $\ \ \ \  0.1 \ \ \ \ $ & $ \ \ \ \ 2.45505 \ \ \ \ $ & $\ \ \ \ 0.02764 \ \ \ \ $ &
  $ \ \ \ \ 0.1 \ \ \ \ $ & $ \ \ \ \ 2.52634 \ \ \ \ $ & $\ \ \ \ 0.04558 \ \ \ \ $\\
  \hline
  $\ \ \ \  0.5 \ \ \ \ $ & $ \ \ \ \ 2.47003 \ \ \ \ $ & $\ \ \ \ 0.02680 \ \ \ \ $ &
  $ \ \ \ \ 0.5 \ \ \ \ $ & $ \ \ \ \ 2.53310 \ \ \ \ $ & $\ \ \ \ 0.04231 \ \ \ \ $\\
  \hline
  $\ \ \ \  1.0 \ \ \ \ $ & $ \ \ \ \ 2.48988 \ \ \ \ $ & $\ \ \ \ 0.02596 \ \ \ \ $ &
  $ \ \ \ \ 1.0 \ \ \ \ $ & $ \ \ \ \ 2.54517 \ \ \ \ $ & $\ \ \ \ 0.03924 \ \ \ \ $\\
  \hline
  \hline
  \multicolumn{3}{c||}{$N_c=3$, $\alpha=-0.001$, $b=-2.82\times10^{-4}$} & \multicolumn{3}{c}{$N_c=3$, $\alpha=0.001$, $b=-2.82\times10^{-4}$}\\
  \hline
  $\beta\times10^3$ & $\mu_c/r_+$ & $T_c/\mu_c$ & $\beta\times10^3$ & $\mu_c/r_+$ & $T_c/\mu_c$ \\
  \hline
  $\ \ \ \  -1.0 \ \ \ \ $ & $ \ \ \ \ 2.59322 \ \ \ \ $ & $\ \ \ \ 0.03756 \ \ \ \ $ &
  $ \ \ \ \ -1.0 \ \ \ \ $ & $ \ \ \ \ 2.59304 \ \ \ \ $ & $\ \ \ \ 0.03754 \ \ \ \ $\\
  \hline
  $\ \ \ \  -0.5 \ \ \ \ $ & $ \ \ \ \ 2.56251 \ \ \ \ $ & $\ \ \ \ 0.02639 \ \ \ \ $ &
  $ \ \ \ \ -0.5 \ \ \ \ $ & $ \ \ \ \ 2.56081 \ \ \ \ $ & $\ \ \ \ 0.02616 \ \ \ \ $\\
  \hline
  $\ \ \ \  -0.1 \ \ \ \ $ & $ \ \ \ \ 2.56685 \ \ \ \ $ & $\ \ \ \ 0.02350 \ \ \ \ $ &
  $ \ \ \ \ -0.1 \ \ \ \ $ & $ \ \ \ \ 2.56566 \ \ \ \ $ & $\ \ \ \ 0.02312 \ \ \ \ $\\
  \hline
  $\ \ \ \  0.1 \ \ \ \ $ & $ \ \ \ \ 2.57202 \ \ \ \ $ & $\ \ \ \ 0.02219 \ \ \ \ $ &
  $ \ \ \ \ 0.1 \ \ \ \ $ & $ \ \ \ \ 2.57097 \ \ \ \ $ & $\ \ \ \ 0.02208 \ \ \ \ $\\
  \hline
  $\ \ \ \  0.5 \ \ \ \ $ & $ \ \ \ \ 2.58557 \ \ \ \ $ & $\ \ \ \ 0.02057 \ \ \ \ $ &
  $ \ \ \ \ 0.5 \ \ \ \ $ & $ \ \ \ \ 2.58472 \ \ \ \ $ & $\ \ \ \ 0.02048 \ \ \ \ $\\
  \hline
  $\ \ \ \  1.0 \ \ \ \ $ & $ \ \ \ \ 2.60638 \ \ \ \ $ & $\ \ \ \ 0.01911 \ \ \ \ $ &
  $ \ \ \ \ 1.0 \ \ \ \ $ & $ \ \ \ \ 2.60568 \ \ \ \ $ & $\ \ \ \ 0.01905 \ \ \ \ $\\
  \hline
  \hline
\end{tabular}
\caption{The numerical values for $\mu_c/r_+$ and $T_c/\mu_c$ with various values of $\beta$.} \label{Nc23-mod}
\end{table}
This table implies that there is actually the range of the parameter $\beta$ for the non-minimal coupled Maxwell field whose presence as the small correction still leads to the CSC phase transition. In addition, we find that the slope $T_c/\mu_c$ of the critical line $T_c=T_c(\mu_c)$ increases as the parameter $\beta$ decreases. This means that compared to the CSC phase explored in the framework of EGB gravity coupled minimally to the vector and scalar fields with including the higher derivative correction for the Maxwell electrodynamics, the presence of the non-minimal coupled Maxwell field with the negative $\beta$ leads to the larger region of the CSC phase and thus it makes the CSC phase more stable. This suggests that
including the non-minimal coupled Maxwell field with the negative $\beta$ would make the gravitational attraction around the event horizon weaker and hence the electrostatic repulsion is easier to overcome the gravitational attraction in order to form the scalar hair.
\section{\label{conclu} Conclusion}

The quark matter at sufficiently high chemical potential and low temperature is expected to exhibit a color superconductivity (CSC) phase which might be present in the cores of neutron stars. In Ref. \cite{Ghoroku2019}, a bottom-up holographic model was introduced in the framework of Einstein gravity to describe the CSC phase in the Yang-Mills (YM) theory. Based on the analysis where the backreaction of the matter part is considered and thus improves the results of the probe approximation, the authors have concluded that the CSC phase appears in the deconfinement phase but not the confinement one for the color number $N_c=1$. However, for $N_c\geq2$ which belongs in the region of the reasonable values of the YM theory, the CSC phase does not appear in both the confinement and deconfinement phases.

Motivated by the above restriction, we have constructed a more realistic holographic model of the CSC phase in the YM theory, which allows to study the CSC phase for the color number $N_c\geq2$. We consider a gravitational system with the matter content consisting of a $U(1)$ gauge field and a charged scalar field in the framework of Einstein-Gauss-Bonnet (EGB) gravity. Here, the gauge field and scalar field are dual to the current of the baryon number and the diquark operator in the boundary field theory, respectively. We have indicated that the Gauss-Bonnet (GB) term plays a role in the breakdown of the Breitenlohner-Freedman (BF) bound and thus the scalar field condensate, corresponding to the occurrence of the CSC phase.

Near the critical chemical potential, the scalar field condensate approaches zero and hence its backreaction on the spacetime geometry is negligible. As a result, the bulk background configuration is given by the EGB gravity coupled to the $U(1)$ gauge field in the asymptotic AdS spacetime. The deconfinement and confinement phases are dual to the planar GB-RN-AdS black hole and GB-AdS soliton, respectively. We have calculated their free energy in the canonical ensemble and obtain the corresponding phase diagram.

When taking the scalar field into account, we study the scalar field condensate and its modification on the phase structure of the bulk background configuration in both the deconfinement and the confinement phases. We determine the necessary condition for destabilizing the scalar field and making it condensing by examining the breakdown of the BF bound. In the deconfinement phase, we found that the CSC phase for $N_c\geq2$ can not be realized in the EGB gravity with the GB coupling parameter $\alpha>0$. However, with $\alpha<0$, the CSC phase is possibly observed for $N_c\geq2$. We solve numerically the equations of motion for the gravitational system in the deconfinement phase and then determine the numerical dependence of the critical chemical potential $\mu_c$ and the critical line $T_c=T_c(\mu_c)$ in terms of the GB coupling parameter. We observe that as increasing the magnitude of the GB coupling parameter the critical chemical potential $\mu_c$ decreases but the slope of the critical line grows corresponding to that the region of the CSC phase is larger. This implies that the scalar field condensate gets easier to form and the CSC phase is more stable in the EGB gravity with the larger magnitude of the GB coupling parameter. Furthermore, by examining the breakdown of the BF bound and solving numerically the equations of motion for the gravitational system in the confinement phase, we show that there is no the CSC phase for the magnitude of $\alpha$ below a certain value which beyond that value it might suggest a breakdown region of the EGB gravity in investigating the CSC phase.

Unfortunately, the occurrence of the CSC phase transition with $N_c\geq2$ within the framework of the EGB gravity requires the magnitude of the GB coupling parameter which is rather large and hence the GB term would no longer be considered as the correction as well as it violates the boundary causality bound. In order to resolve this problem, we include additionally the higher derivative correction for the Maxwell electrodynamics and the non-minimal coupled Maxwell field. We show the existence of the values for the parameters characterizing the higher derivative corrections under consideration where they are the small corrections and work to realize the CSC phase transition with $N_c\geq2$. In comparison with Einstein gravity, the presence of the non-minimal coupled Maxwell field and the higher derivative correction for the Maxwell electrodynamics with the negative characteristic parameters yields the weaker gravitational attraction around the event horizon in analogy to the EGB gravity of the negative GB coupling parameter but with the stronger effect. Thus, the electrostatic repulsion is easier to overcome the gravitational attraction resulting in the condensation of the scalar field even the small magnitude of the GB coupling parameter. 

\section*{Acknowledgments}
We would like to express sincere gratitude to Referees for their constructive comments, suggestions and questions by which the quality of the paper has been improved.

\end{document}